\documentclass[iop]{emulateapj}

\usepackage{apjfonts}
\usepackage{CJK}
\usepackage{color}
\usepackage{soul}
\usepackage{hyperref}

\shorttitle{PROSPECTS FOR CHEMICAL TAGGING}
\shortauthors{TING, CONROY, \& GOODMAN}

\begin{document}

\begin{CJK*}{UTF8}{gbsn}
\title{Prospects for Chemically Tagging Stars in the Galaxy}
\author{Yuan-Sen Ting (丁源森), Charlie Conroy, Alyssa Goodman}
\affil{Harvard--Smithsonian Center for Astrophysics, 60 Garden Street, Cambridge, MA 02138, USA}

\slugcomment{Submitted to ApJ}

\begin{abstract} 

It is now well-established that the elemental abundance patterns of stars holds key clues not only to their formation but also to the assembly histories of galaxies. One of the most exciting possibilities is the use of stellar abundance patterns as ``chemical tags'' to identify stars that were born in the same molecular cloud. In this paper we assess the prospects of chemical tagging as a function of several key underlying parameters. We show that in the fiducial case of $10^4$ distinct cells in chemical space and $10^5-10^6$ stars in the survey, one can expect to detect $\sim10^2-10^3$ groups that are $\ge5\sigma$ overdensities in the chemical space. However, we find that even very large overdensities in chemical space do not guarantee that the overdensity is due to a single set of stars from a common birth cloud. In fact, for our fiducial model parameters, the typical $5\sigma$ overdensity is comprised of stars from a wide range of clusters with the most dominant cluster contributing only 25\% of the stars. The most important factors limiting the identification of disrupted clusters via chemical tagging are the number of chemical cells in the chemical space and the survey sampling rate of the underlying stellar population. Both of these factors can be improved through strategic observational plans. While recovering individual clusters through chemical tagging may prove challenging, we show, in agreement with previous work, that different CMFs imprint different degrees of clumpiness in chemical space. These differences provide the opportunity to statistically reconstruct the slope and high mass cutoff of CMF and its evolution through cosmic time.  

\end{abstract}


%
%
%
%
%
%
\section{Introduction}
\label{sec:introduction}

Despite decades of effort, we still lack a thorough understanding of how galaxies assemble and evolve over cosmic time. This is true not only for distant galaxies but also for our own Milky Way. In the current paradigm, galaxies such as the Milky Way form from smaller pieces \citep[e.g.,][]{sea78}, driven by the hierarchical growth of dark matter structures \citep[e.g.,][]{pee71, pre74}. Much of the most exciting phases of star formation and galaxy assembly appear to have taken place at early times, perhaps before $z\sim2$. If true, this puts much of the most interesting phases of galaxy formation beyond direct detailed study. For this reason much effort has focused on reconstructing the past based on present-day observations of stars, in particular in the Galaxy. For example, studies of the Galactic stellar halo provides clues to the assembly history of dwarf galaxies \citep[e.g.][]{egg62, sea78}. The properties of stars in the thin and thick disks provide clues to the formation history of these Galactic components. The abundance patterns of the most metal poor stars probe star formation and supernovae conditions during the first generation of stars. And the evolutionary histories of star clusters, both intact, dissolving, and long destroyed, offer clues not only into the star formation process (by reconstructing the CMF), but also the dynamical history of the Galaxy \citep[e.g.,][]{kol07,all12,web13}.

However, reconstructing disrupted star clusters is difficult because most star clusters dissolve quickly upon their formation due to dynamical interactions, such as intracluster $N$-body interaction and external tidal stripping from ram pressure. In fact, most clusters are not expected to survive for more than 10 Myrs \citep{lad03}. For this reason, the study of young embedded clusters \citep[e.g.,][]{bic03,por03,kop08,bor11} is typically restricted to the study of star formation conditions at the present time. Although most star clusters are quickly disrupted, they retain their identity in kinematic phase space for a longer period of time. Several examples of clusters identified in phase space are known, such as HR1614, the Argus association and the Wolf 360 group \citep[e.g.,][]{des07a,des13,bub10}, with an age of 2-3 Gyrs. This implies that at least some clusters can maintain their phase space identity for a few disk dynamical times. Within a few dynamical times these groups will phase mix with the background stars, which implies that the timescale over which groups can be identified in phase space is still a small fraction of the age of the Galaxy.

While dynamical information is mostly short-lived, elemental abundances are expected to leave a more permanent fossil record of star clusters. The idea of ``chemical tagging'', first proposed by \citet{fre02} \citep[also see][]{bla14}, is to use elemental abundances to identify stars that are now widely separated in phase space to a common birth site. If such an association could be made, even for a small fraction of stars, it would provide an extraordinary new view into both the early star formation process and the subsequent dynamical history of the Galaxy.

Observations have shown that satellite galaxies exhibit different chemical evolution histories compared to stars either in the disk, bulge, or halo of the Galaxy \citep[e.g.,][]{ven04, pom08, ven08, tol09, let10}. As a consequence, stars accreted into the Galaxy from different satellite systems should show distinct chemistry from e.g., disk stars. It has been proposed that these variations could be used in chemical tagging to find the remnants of disrupted satellite galaxies \citep{fre02}. The possibility of reconstructing disrupted satellite galaxies via chemical tagging could for example provide important clues to the missing satellite problem \citep{moo99}.

Previous studies of high-resolution stellar spectroscopy were limited to a few hundred stars \citep[e.g.,][]{bar05, red06, ben14}. The small samples restricted the possibility of chemical tagging for reasons that will become clear in later sections. But this situation is rapidly changing. Recent and on-going large-scale surveys, such as GALAH \citep{des15}, Gaia-ESO \citep{ran13} and APOGEE \citep{zas13} aim to observe $10^5-10^6$ stars with resolution $R > 20,000$ in order to measure $\sim 15 - 30$ elements for each star. These surveys were motivated, at least in part, by the idea of chemical tagging and the prospects for uncovering the distribution of stars in their $N-$dimensional chemical space, spanned by the elemental abundances.

There are several conditions that must be met for chemical tagging to work \citep[see][for details]{fre02, bla04, bla10a, bla10b, des15}. First, clusters must be internally chemically homogeneous. Open clusters have been found to be chemically homogeneous at the level of $\sigma_{[X/{\rm Fe}]} < 0.05$ dex \citep[e.g.,][]{des07b, des09, tin12b, fri14, one14}. Theoretical arguments from \citet{bla10b} showed that the chemical signature within a protocloud should have sufficient time to homogenize before the first supernova goes off, for clusters with mass $10^5 - 10^7 \, M_\odot$. Simulations by \citet{fen14} showed that turbulent mixing, even for a loosely bound cluster, could homogenize the elemental abundances of a protocloud. Their simulations showed that turbulent mixing creates an intracluster chemical dispersion at least five times more homogenized than the protocloud. Both observations and theory agree that clusters less massive than $\sim10^7 \, M_\odot$ should be chemical homogeneous, except perhaps for the confounding internal abundance trends observed in the light elements of all known globular clusters \citep[e.g.,][]{car09,mar11}, though many globular clusters show a high degree of chemical uniformity \citep[e.g.,][]{roe15} in all heavy elements.

In addition to cluster homogeneity, the existence of substantial cloud-to-cloud variation in elemental abundances is another requirement. For example, if all star clusters shared the same elemental abundances, it would not be possible to separate them in chemical space. We know that this condition is broadly satisfied given the sizable spread in abundance ratios in existing spectroscopic samples \citep[e.g.,][]{edv93,ben14}. Quantitatively, an important parameter is the volume of abundance space that is available for a particular survey. This volume depends both on Galactic chemical evolution and on the particular survey design. The latter is important both in determining the target sample and in the number of elements that can be spectroscopically measured. Combining the available chemical volume with the measurement uncertainty on individual abundances allows us to define the concept of the total number of distinct cells in chemical space. As we will see below, this is a key concept in chemical tagging \citep[see also][]{fre02}.

\citet{tin12a} presented an empirical estimate of cloud-to-cloud variation in elemental abundances. They performed principal component analysis and estimated that there are $7-9$ independent dimensions among the $\sim 25$ elements that will be measured by surveys such as GALAH and Gaia-ESO, and $4-5$ independent dimensions for an APOGEE-like survey. From this one can estimate the number of distinguishable cloud-to-cloud variations in the chemical space, denoted $N_{\rm cells}$. As discussed in detail in \S\ref{subsec:chem-model} below, the result is that modern surveys should be able to reach $N_{\rm cells} \sim 10^{3-4}$, at least, implying that there is a decent cloud-to-cloud variation.

The goal of this paper is to explore the prospects for identifying long disrupted star clusters based on their clustering in chemical space. We follow \citet{fre02}, \citet{bla10a}, and \citet{des15} in identifying the global survey parameters and the shape of the CMF as key parameters. Our emphasis on the information contained in the distribution (i.e., clumpiness) of stars in chemical space echoes the results found in \citet{bla10a}. In the present work we consider a wide array of parameters in order to identify optimal regions of parameter space for chemical tagging. In addition, for the first time we analyze the local properties of cells in chemical space that appear as high sigma fluctuations and find that in many cases these high overdensities in chemical space are not the result of a single star cluster but instead are comprised of stars from many distinct birth sites.

The rest of this paper is organized as follows. In \S\ref{sec:overview} we review several basic arguments relevant for chemical tagging and in \S\ref{sec:models} we describe the model used in the present work. In \S\ref{sec:results} we present the results and discuss how these assumptions and survey strategies affect the chemical tagging detections. In \S\ref{sec:discussion} we discuss various caveats, limitations and future directions. We conclude in \S\ref{sec:conclusions}. It is difficult to present the full set of results from a multidimensional parameter space and so we urge readers to explore the online interactive applet \footnote{\href{www.cfa.harvard.edu/~yuan-sen.ting/chemical_tagging.html}{www.cfa.harvard.edu/$\sim$yuan-sen.ting/chemical$\_$tagging.html}} created in the course of this project (see Appendix~\ref{sec:interactive} for details).

\begin{figure*}
\center
\includegraphics[width=\textwidth,natwidth=1700,natheight=900]{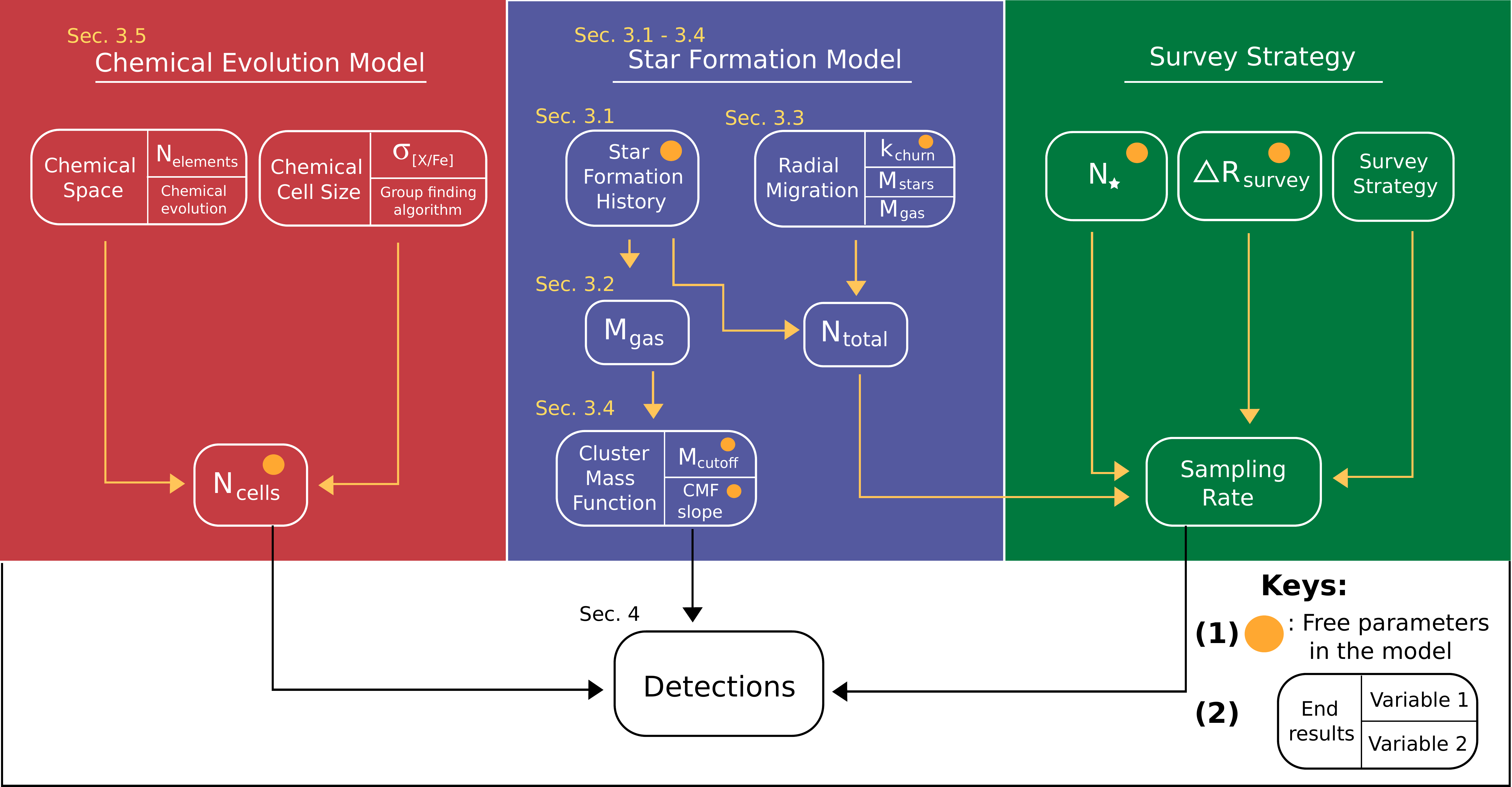}
\caption{Flow chart demonstrating the main components of the model. Sections defining or describing certain components of the model are indicated in the chart.}
\label{fig:flow-chart}
\end{figure*}

%
%
%
%
%
%

\section{Basic Arguments}
\label{sec:overview}

As we will show quantitatively below, the prospects for chemical tagging largely depends on the number of stars sampled per cluster. This number in turn primarily depends on the number of stars in the survey divided by the integrated star formation rate (SFR), over cosmic history, in the volume sampled by the survey. We will denote the former number as $N_\star$, the latter number as $M_{\rm annulus}$. Ongoing and upcoming surveys are targeting primarily FGK stars, which have on average $\langle M \rangle \approx 1 M_\odot$. This implies that $N_\star$ stars in a survey corresponds to $N_\star$ in solar masses and therefore numerically $M_{\rm annulus}\approx N_{\rm annulus}$. The ratio of $N_\star$ and $N_{\rm annulus}$ defines the sampling rate. In this section, we motivate why the sampling rate largely defines the number of stars sampled per cluster \citep[see also][]{des15}.

First, let's consider a simple case where there is no radial migration and stellar excursion, i.e., stars stay in the annulus in which they were born. The integrated SFR in the Solar annulus, with a survey width $\Delta R_{\rm survey} = \pm 3 \, {\rm kpc}$, is $\sim 2 \times 10^{10} \, M_\odot$ (see model detail in \S\ref{sec:models}).\footnote{The survey width $\Delta R_{\rm survey}$ defines the Solar annulus by $|R-R_0| < |\Delta R_{\rm survey}|$. The survey width should not be confused with the line-of-sight depth from the Sun, which is $|{\bf R} - {\bf R_0}| < 3 \, {\rm kpc}$.} For a survey of $10^6$ stars with $\langle M \rangle = 1 \, M_\odot$, the sampling rate can thus be calculated to be $(10^6 \, M_\odot)/(2 \times 10^{10} \, M_\odot) = 1/(2 \times 10^{4})$. In other words, assuming all stellar mass (including stellar mass loss) is now fully mixed in the annulus, we would have only sampled, on average, $1/(2 \times 10^{4})$ of the original zero age mass from each cluster. Thus, we would expect to observe, on average, only one star from a $2 \times 10^4 \, M_\odot$ cluster. If we define the ``detection'' of a cluster to include the identification of at least 10 stars, then for a survey of $10^6$ random stars in the solar annulus we would be able to probe clusters more massive than $2 \times 10^5 \, M_\odot$.

In practice, the sample is affected by the process of radial migration \citep[e.g.,][]{bla10b}. Some stars are migrated away from their birth annulus while others that were born outside the Solar annulus will now reside within the Solar annulus. In other words, the number of stars that could end up in the Solar annulus increases with radial migration (another way of thinking of this effect is that the effective volume of the Solar annulus increases as the strength of radial migration increases). Given that the number of stars in the survey stays the same, the sampling rate decreases with radial migration. For a fixed survey strategy, the minimum cluster mass that one can probe increases in the presence of radial migration. 

We must also consider the fact that we have limited resolution in separating groups in terms of their elemental abundance variations due to measurement uncertainties on the abundances. Multiple clusters might share the same cell in chemical space \citep[e.g.,][]{bla10a}. If we assume a CMF over the range $50 \, M_\odot$ to $10^6 \, M_\odot$ and a CMF slope of $-2$ (see details in \S\ref{sec:models}), the mean cluster mass is $\sim 5 \times 10^2 \, M_\odot$. Since the integrated SFR is $\sim 2 \times 10^{10} \, M_\odot$, we deduce that there are $\sim 4 \times 10^7$ clusters in the Solar annulus. Fully resolving clusters in chemical space would require roughly as many distinct chemical cells \citep{fre02}, but it was argued in the Introduction that the actual number of chemical cells spanned by the data may be $2-3$ orders of magnitude lower. This suggests that most cells in chemical space will be occupied by many clusters, each with a small number of stars sampled per cluster. One of the key goals of this paper is to understand the distribution of clusters in chemical space under different scenarios.

The simple calculations in this section already demonstrate that key parameters include the number of stars in a survey, $N_\star$, the geometry of the survey (via $M_{\rm annulus}$), the strength of radial migration, the shape of the CMF (which sets the typical cluster size), and the number of cells in chemical space ($N_{\rm cells}$).
\begin{table*}
\begin{center}
\caption{List of constraints in this study.\label{table:constraints}}
\begin{tabular}{lll}
\tableline \tableline
\\[-0.2cm]
Property                                           & Value                                               & References \\[0.1cm]
\tableline
\\[-0.2cm]
Galactocentric radius of the Sun, $R_0$            & $8 \, {\rm kpc}$                                    & \citet{ghe08,gil09,rei14} \\[0.1cm]
Stellar surface density, $\Sigma_\star (R_0,z=0)$  & $38 \, M_\odot {\rm pc}^{-2}$                       & \citet{fly06,bov13,zha13} \\[0.1cm]
Gas surface density, $\Sigma_{\rm gas} (R_0,z=0)$  & $13 \, M_\odot {\rm pc}^{-2}$                       & \citet{fly06} \\[0.1cm]
Total stellar mass in the disk, $M_\star (z=0)$    & $4.5 \times 10^{10} \, M_\odot$                     & \citet{fly06,bin08,bov13} \\[0.1cm]
Halo virial mass, $M_{\rm halo} (z=0)$             & $10^{12} \, M_\odot$                                & \citet{wil99,kly02,xue08,kaf12} \\[0.1cm]
Global SFR ($z=0$)                                 & $0.5 - 2 \, M_\odot {\rm yr}^{-1}$                  & \citet{rob10,cho11,ven13} \\[0.1cm]
Solar neighborhood SFR, $\Sigma_{\rm SFR} (R_0,t)$ & $3-6 \, M_\odot {\rm Gyr}^{-1} {\rm pc}^{-2}$       & \citet{her00,ber01} \\[0.1cm]
Stellar disk scale length, $R_\star (z=0)$         & $2.2 \, {\rm kpc}$                                  & \citet{bov13} \\[0.1cm]
SFR scale length, $R_{\rm SFR} (z=0)$              & $2.6 \, {\rm kpc}$                                  & \citet{sch11} on NGC 6946 \\[0.1cm]
Gas scale length, $R_{\rm gas} (z=0)$              & $4.2 \, {\rm kpc}$                                  & \citet{sch11} on NGC 6946 \\[0.1cm]
Radial size growth                                 & $R_\star \propto M_\star^{0.27}$                    & \citet{van13} \\[0.1cm]
\tableline
\end{tabular}
\end{center}
\end{table*}
\begin{table}
\begin{center}
\caption{List of parameters in the model.\label{table:parameters}}
\begin{tabular}{llll}
\tableline \tableline
\\[-0.2cm]
Parameter                                                                        & Fiducial                              & Range \\[0.1cm]
\tableline
\\[-0.2cm]
In-situ fraction, $f_{\rm in-situ} (\Delta R_{\rm survey} = \pm 1 \, {\rm kpc})$ & $50\%$                                & $15\%-100\%$ \\[0.1cm]
Survey width, $\Delta R_{\rm survey}$                                            & $\pm 3 \, {\rm kpc}$                  & $\pm 0.6 - 5 \, {\rm kpc}$ \\[0.1cm]
CMF slope, $\alpha$                                                              & $-2.0$                                & $-1.5$ to $-2.5$ \\[0.1cm]
CMF low mass cutoff, $M_{\rm cluster}^{\rm min}$                                 & $50 \, M_\odot$                       & $10-100 \, M_\odot$ \\[0.1cm]
CMF high mass cutoff, $M_{\rm cluster}^{\rm max}$                                & see Figure \ref{fig:model-properties} & see Figure \ref{fig:model-properties} \\[0.1cm]
Number of chemical cells, $N_{\rm cells}$                                        & $10^4$                                & $10^3-10^5$ \\[0.1cm]
Number of stars in the survey, $N_\star$                                         & $10^6$                                & $10^4-10^6$ \\[0.1cm]
\tableline
\end{tabular}
\end{center}
\end{table}

\begin{table*}
\begin{center}
\caption{Meaning of other important symbols in this paper that are not listed in Table~\ref{table:constraints} and \ref{table:parameters}.\label{table:other-symbols}}
\begin{tabular}{llll}
\tableline \tableline
\\[-0.2cm]
Symbols               & Meanings \\[0.1cm]
\tableline
\\[-0.2cm]
$k_{\rm ch}$          & Churning strength in the radial migration prescription \\[0.1cm]
$\eta$                & Gas fraction; the ratio of gas mass over total dynamical mass \\[0.1cm]
$\sigma_{\rm [X/Fe]}$ & Elemental measurement uncertainty in [X/Fe] \\[0.1cm]
$\sigma$              & Elemental measurement uncertainty along the chemical space principal components \\[0.1cm]
$N_{\rm dim}$         & Number of independent/informative dimensions in chemical space \\[0.1cm]
$M_{\rm gas}$         & Total gas mass in the Milky Way \\[0.1cm]
$M_{\rm cluster}$     & Zero age stellar mass of a star cluster \\[0.1cm]
$M_{\rm annulus}$     & Integrated SFR, over cosmic history, in the volume sampled by the survey \\[0.1cm]
$N_{\rm annulus}$     & Total number of stars (including stellar mass loss) in the volume sampled by the survey \\[0.1cm]
$N_i$                 & Total number of stars sampled in a chemical cell \\[0.1cm]
$N_{\rm mean}$        & Average number of stars sampled per chemical cell \\[0.1cm]
$N_{\rm cluster}$     & Number of stars sampled from a cluster \\[0.1cm]
$N_{\rm dominant}$    & Number of stars sampled from the most dominant cluster in a chemical cell \\[0.1cm]
local S/N             & Number of stars sampled from the most dominant cluster over the total number of other stars in a chemical cell \\[0.1cm]
$f_{\rm sub}$         & Sampling rate of a certain stellar subpopulation \\[0.1cm]
\tableline
\end{tabular}
\end{center}
\end{table*}

%
%
%
%
%
%
\section{Model Description}
\label{sec:models}

In this section we describe the ingredients of our model for the Milky Way in some detail. The model is spatially two dimensional (though we assume that stars are uniformly distributed in the azimuthal angle), time-dependent, and statistical in nature. For the present study we are only interested in the disk; the bulge and halo are not included in the model below. We do not follow dynamics nor do we include a treatment of chemical evolution (these will be subjects of future work). The present aim is to build a model that is computationally very fast to allow the exploration of a large multi-dimensional parameter space.

The model specifies the star formation history (SFH) and evolution in time of the size of the Milky Way disk and the gas mass distribution. We define the SFH to be the total SFR in the Milky Way as a function of cosmic time. These quantities are used to model the effects of radial migration and an evolution in the cutoff of the CMF. The model is illustrated in a flow chart in Figure \ref{fig:flow-chart}. Table \ref{table:constraints} lists observational constraints that we employ to constrain the model. Free parameters in the model and their adopted fiducial values are listed in Table \ref{table:parameters}. We now proceed to explain the details of the model.

%
%
%
%
%
%
\subsection{Star formation history and radial size growth of the disk}
\label{subsec:sfh}

The SFH in the Solar neighborhood, $\Sigma_{\rm SFR} (R_0,t)$, has been estimated by analyzing the color-magnitude diagram from the Hipparcos catalog. Results from, for e.g., \citet{her00} and \citet{ber01} showed a rather flat SFH near $R_0$, ranging from $3 - 6 \, M_\odot {\rm Gyr}^{-1} {\rm pc}^{-2}$ through $0-8$ Gyr in lookback time. The current total SFR in the Milky Way has been estimated to be $0.5 - 2 \, M_\odot {\rm yr}^{-1}$ from the study of young stellar objects \citep[e.g.,][]{rob10,cho11,ven13}. 

In comparison to the Solar neighborhood, the Galactic global SFH is less well understood. We therefore adopt cosmological semi-empirical modeling from \citet{beh13}, assuming a Milky Way halo virial mass of $M_{\rm halo} \equiv M_{200} = 10^{12} \, M_\odot$ \citep[e.g.,][]{wil99, kly02, xue08, kaf12}. \citet{beh13} investigated the best-fitting global SFH as a function halo mass that is consistent with the observed galaxy stellar mass function, specific SFR, and cosmic SFR. We fit their result for Milky Way-like halos with a Schechter function, 
\noindent
\begin{equation} 
{\rm SFR} [M_\odot {\rm yr}^{-1}] = A \, (t[{\rm Gyr}]/C)^B \, \exp( - t[{\rm Gyr}]/C).
\end{equation}

Given a global SFH, the stellar mass evolution is calculated assuming the stellar population synthesis code from \citet{con09}, with a Kroupa IMF \citep{kro02} from $0.08 - 125 \, M_\odot$. The synthesis code is used to take into account secular stellar mass loss, etc. The normalization of the global SFH is further adjusted such that the present-day stellar mass (long-lived stars + remnant stars) agrees with observations, $M_\star (z=0) = 4.5 \times 10^{10} \, M_\odot$ \citep[e.g.,][]{bin08, bov13}. In this study, we only trace long-lived stars with $0.5 - 1.5 \, M_\odot$ because almost all FGK stars in chemical tagging surveys are within this mass range.

\begin{figure*}
\center
\includegraphics[width=\textwidth,natwidth=1400,natheight=1000]{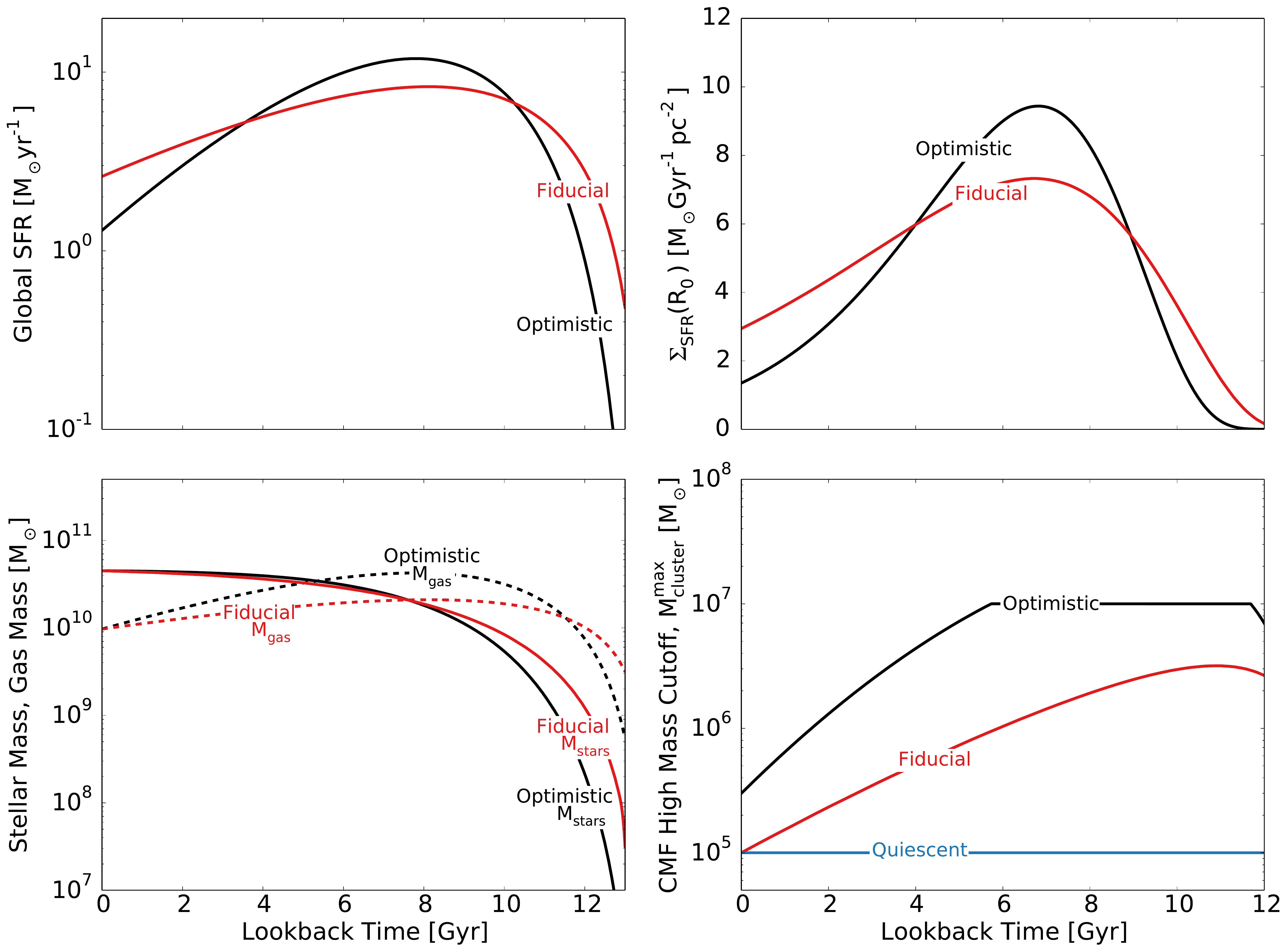}
\caption{{\em Bottom right panel}: Evolution of CMF high mass cutoff. The CMF evolves according to \citet{esc08}. The CMF cutoff is the main property that defines the quiescent, fiducial and optimistic models that we will discuss throughout this study. For example, the optimistic CMF allows the formation of larger clusters ($M_{\rm cluster} \sim 10^7 \, M_\odot$). We adopt an upper limit of $M_{\rm cluster}^{\rm max} = 10^7 \, M_\odot$, above which clusters are not expected to be homogeneous. {\em Bottom left panel}: Stellar and gas mass evolutions. The gas mass at $z=0$ is calculated from $\Sigma_{\rm gas} (R_0,z=0) = 13 \, M_\odot {\rm pc}^{-2}$. The gas mass evolution is calculated from the global SFH, following a Kennicutt-Schmidt law with $\alpha_{\rm KS} = 1.5$. {\em Top left panel}: Global SFH models in this study, assuming $M_{\rm halo}=10^{12} \, M_\odot$ adjusted to produce $M_\star (z=0) = 4.5 \times 10^{10} \, M_\odot$. The two SFHs have the same integrated SFR. The SFHs mainly come into play in determining the gas mass evolution and subsequently the CMF cutoff evolution. Since the quiescent CMF cutoff is constant through cosmic time without evolution, employing the optimistic SFH or fiducial SFH for the quiescent model does not change its results as they have the same integrated SFR. We choose to follow the fiducial SFH for the quiescent model as it fits the $\Sigma_{\rm SFR} (R_0,t)$ better. {\em Top right panel}: $\Sigma_{\rm SFR} (R_0,t)$ calculated from the global SFHs.}
\label{fig:model-properties}
\end{figure*}

We consider two SFH models in this study, with parameters from equation (1) as follows: (1) $A=1.4$, $B=4.4$, $C=1.3$, which is the best fitting SFH model from Behroozi et al.; (2) $A=15.5$, $B=2$, $C=2.7$, which produces better agreement with the observed $\Sigma_{\rm SFR} (R_0,t)$. Both models are within the uncertainty quoted by Behroozi et al. We adopt the latter as the fiducial model and the former to be the optimistic model (see Figure~\ref{fig:model-properties} and Table~\ref{table:models}). The former coins the term ``optimistic model'' as its more highly peaked SFR entails a higher total gas mass (see \S\ref{subsec:gas-mass}). The higher total gas mass in turn predicts a larger cluster high mass cutoff (see \S\ref{subsec:cmf}) than the ``fiducial model.'' We emphasize that while the optimistic and fiducial models assume different SFHs, the integrated SFRs of these models over cosmic time are the same. Since the total integrated SFRs are the same, they both produce the same $M_\star(z=0)$ and $\Sigma_\star (R_0,z=0)$. Therefore, the sampling rate is the same for both cases. The global SFR and $\Sigma_{\rm SFR} (R_0,t)$ in these two models are compared in the upper panels in Figure \ref{fig:model-properties}. The main differences of these models are summarized in Table~\ref{table:models} (the ``quiescent model'' will be defined in \S\ref{subsec:cmf}).

With the stellar mass evolution in hand, we then derive the radial size growth of the Milky Way using the empirical relation from \citet{van13}. By studying the evolution of galaxies at a fixed comoving number density at different redshifts, \citet{van13} found that the effective radius $R_\star$ of Milky Way-like galaxies grow with the total stellar mass according to the relation $R_\star \propto M_\star^{0.27}$.

Finally, to fully specify the star formation at different radii, we also require the star formation scale length, $R_{\rm SFR}$, and its evolution. Unfortunately, determining $R_{\rm SFR}$ for the Milky Way is observationally challenging. Therefore, we resort to $R_{\rm SFR}$ from extragalactic studies where the external vantage point provides an easier measurement of scale lengths. NGC 6946 has long been thought to be a Milky Way counterpart \citep[e.g.,][]{ken12}. We find the SFR and the (atomic and molecular) gas mass of NGC 6946 from \citet{sch11} can be fitted with an exponential model. We find scale lengths $R_{\rm SFR}(z=0) = 2.6 \, {\rm kpc}$ and $R_{\rm gas}(z=0) = 4.2 \, {\rm kpc}$, which we adopt in our model of the Milky Way. To compute the evolution $R_{\rm SFR}(z)$ and $R_{\rm gas}(z)$ through cosmic time, we assume all scale lengths trace the stellar effective radius. We find that this adopted $R_{\rm SFR}(z)$ leads to a stellar disk scale length of $R_\star (z=0) = 2.2 \, {\rm kpc}$ and $\Sigma_\star (R_0,z=0) = 38 \, M_\odot {\rm pc}^{-2}$. These values agree with existing observations \citep{fly06, bov13, zha13}. Furthermore, the model implies $R_{\rm gas}(z=0) \simeq 2 R_\star(z=0)$, agreeing with \citet{bov13}.
\begin{table}
\begin{center}
\caption{Summary of the three model variants in this study.\label{table:models}}
\begin{tabular}{llll}
\tableline \tableline
\\[-0.2cm]
Property                   & Optimistic                    & Fiducial                       & Quiescent \\[0.1cm]
\tableline
\\[-0.2cm]
CMF cutoff                 & $\sim 10^7 \, M_\odot$  & $\sim 10^6 \, M_\odot$   & $10^5 \, M_\odot$ \\[0.1cm]
Global SFR                 & Peaks in the past             & More flat                      & More flat \\[0.1cm]
$\Sigma_{\rm SFR} (R_0,t)$ & Too high in the past          & Agrees with obs.               & Agrees with obs. \\[0.1cm]
Integrated SFR             & The same                      & The same                       & The same \\[0.1cm]
\tableline
\end{tabular}
\end{center}
\end{table}

%
%
%
%
%
%
\subsection{Gas mass distribution \& evolution}
\label{subsec:gas-mass}

The mass of gas in the disk comes into play in two aspects of the model, namely the radial migration prescription and the CMF evolution. We assume $\Sigma_{\rm gas} (R_0,z=0) = 13 \, M_\odot {\rm pc}^{-2}$ \citep{fly06}, which, when combined with $R_{\rm gas} (z=0) = 4.2 \, {\rm kpc}$, yields a total gas mass of $M_{\rm gas} (z=0) = 9.7 \times 10^9 \, M_\odot$. We the estimate the redshift evolution of the gas mass $M_{\rm gas} (z)$ by inverting the Kennicutt-Schmidt relation with $\alpha_{\rm KS} = 1.5$ and the SFR evolution described in the previous section. The distribution of gas is fully specified by $M_{\rm gas} (z)$ and $R_{\rm gas}(z)$. The total stellar mass and the total gas mass evolution are shown in the bottom left panel in Figure \ref{fig:model-properties}. For this work we do not need to specify the disk scale height because all quantities of interest are related to surface mass densities.

\begin{figure}
\includegraphics[width=0.45\textwidth]{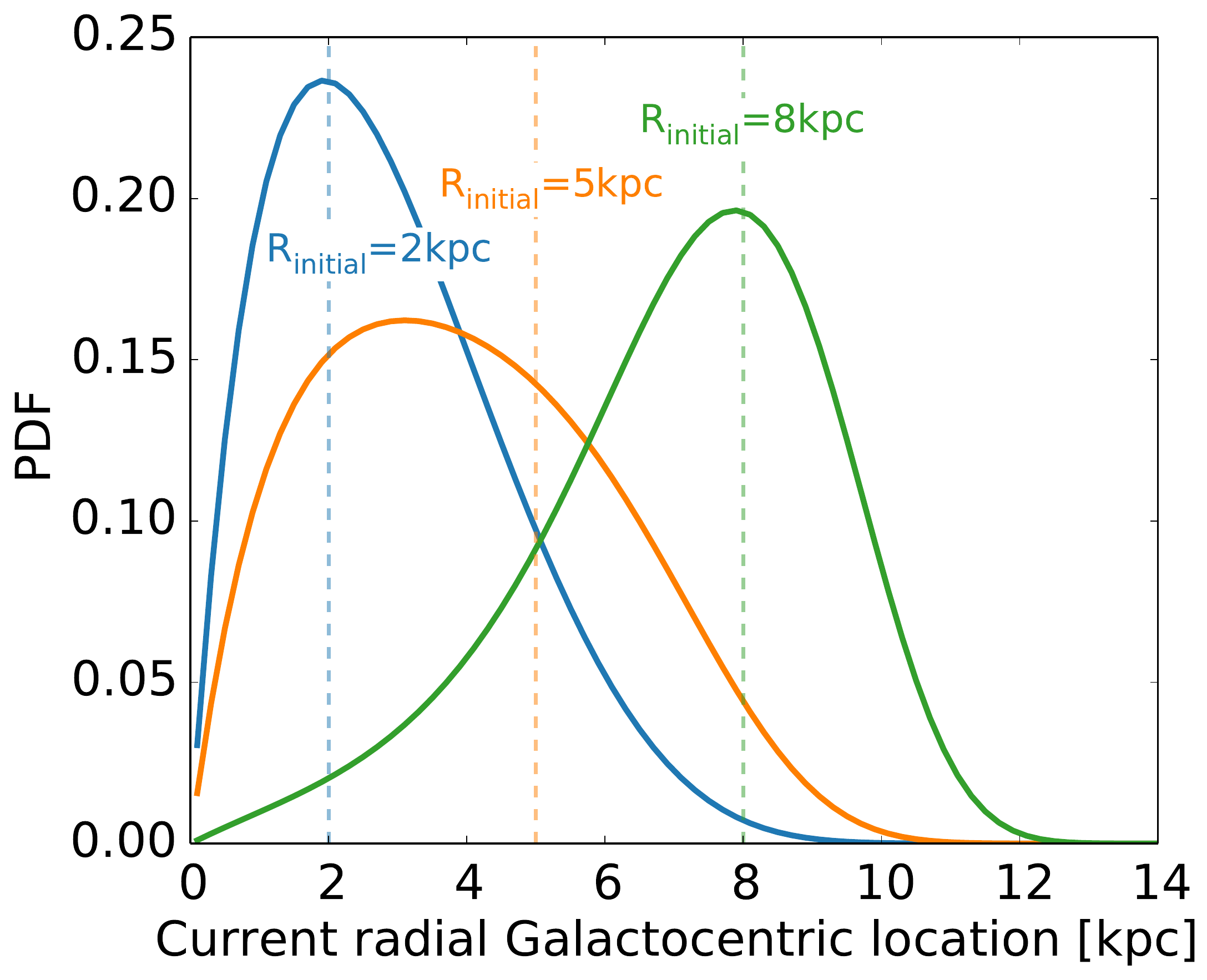}
\caption{Probability of position of a star after evolving over 13 Gyr, assuming $f_{\rm in-situ} = 50\%$. The solid lines show the final positions, whereas the dashed lines show the corresponding initial positions.}
\label{fig:radial-migration}
\end{figure}

%
%
%
%
%
%
\subsection{Radial migration}
\label{subsec:migration}

Radial migration describes the phenomenon of stars in the disk moving, either inward or outward, in radius from their birth radius. Studies of processes giving rise to radial migration have a long history. In the past decade, radial migration has gained increasing attention as playing a key role in driving the chemodynamical evolution of the Milky Way \citep[e.g.,][]{sel02, hay08, sch09, min10, bla10b, dim13}. Due to its role in changing stellar orbiting radii, radial migration provides tentative explanations to some observational puzzles. For example, the upturn in the stellar population age at the outer part of some galaxies \citep[e.g.,][]{bak08, zhe15}, the wide range of stellar metallicity in the Solar neighborhood \citep[e.g.,][]{hay08,sch09}; and perhaps even the formation of the thick disk \citep[e.g.,][]{loe11} can be explained by appealing to the process of radial migration.

An important physical process giving rise to radial migration is known as ``churning'' \citep{sel02}. In the process of churning, stars that co-rotate with transient non-axisymmetric features can increase their angular momentum while maintaining the ellipticity of the orbit, effectively bumping stars from an orbiting radius to the other. \citet{sch09} proposed a simple analytic formula for churning that we will adopt in this study. In this prescription, the probability of moving from the $i$-th to the $j$-th annulus, $P_{ij}$, where $j = i \pm 1$, is given by
\noindent
\begin{equation}
P_{ij} = k_{\rm ch} \frac{M_j}{M_{\rm max}},
\end{equation}

\noindent
where $M_j$ denotes the total (stellar + gas) mass of the $j$-th annulus and $k_{\rm ch}$ is a free parameter governing the strength of the churning.

In the present work we discretize the model galaxy into annuli with width of $0.2 \, {\rm kpc}$ and apply the churning exchange every 0.5 Gyr. We define in-situ fraction, $f_{\rm in-situ}$, as the fraction of stars that were born in-situ in a Solar annulus with $\Delta R_{\rm survey} = \pm 1 \, {\rm kpc}$. Clearly, $f_{\rm in-situ}$ depends on the choice of $\Delta R_{\rm survey}$. We choose $\Delta R_{\rm survey} = \pm 1 \, {\rm kpc}$ to calculate the in-situ fraction, instead of our fiducial value $\pm 3 \, {\rm kpc}$ in the model for ease of comparing to hydrodynamics simulations \citep[e.g.,][]{ros08}. We note that the free parameter $k_{\rm ch}$ maps directly into the variable $f_{\rm in-situ}$, and we choose to express the effect of radial migration in terms of the latter value. We consider a range of $k_{\rm ch}$ corresponding to $f_{\rm in-situ} = 15\%- 100\%$ and we choose $f_{\rm in-situ} = 50 \%$ to be the fiducial value, as suggested by simulations \citep[e.g.,][]{ros08, hal15}. To illustrate the radial migration prescription adopted in this study, solid lines in Figure \ref{fig:radial-migration} show the PDF of the final position of a star after 13 Gyr of evolution starting from various initial positions. 

In addition to churning, scattering, e.g., from interactions with molecular clouds, can also diffuse stars from their birth radii. This scattering is known as ``blurring'' \citep{sel02}. For simplicity, we do not include blurring in the model. However, we note for our purposes only the fraction $f_{\rm in-situ}$ is important; the details of migration, either through churning or blurring are largely irrelevant in this study.

%
%
%
%
%
%
\subsection{Cluster mass function evolution}
\label{subsec:cmf}

We have discussed in \S\ref{sec:overview} that the number of stars sampled {\it per cluster} is governed primarily by the sampling rate and the in-situ fraction. However, knowing the detections per cluster is insufficient. To determine the number of detectable groups, we also need to understand the relative number of massive clusters compared to their smaller counterparts. Therefore, the CMF is another key factor \citep[see also][]{bla10a}. In this study, we assume a CMF that is characterized by a power law slope $\alpha$, high mass cutoff $M_{\rm cluster}^{\rm max}$ and low mass cutoff $M_{\rm cluster}^{\rm min}$, where
\begin{equation} 
\frac{{\rm d} N}{{\rm d} M} \propto M^{-\alpha}.
\end{equation}

\noindent
Note that cluster masses refer to zero age masses; clusters will lose at least a factor of two mass after a Hubble time due to stellar evolution effects and the evaporation of stars.

\citet{lad03} analyzed young embedded clusters within $2.5 \, {\rm kpc}$ from the Sun and found a CMF slope $\alpha \approx -2.0$. We take this as the fiducial value in the model. The fact that $\alpha \approx -2$ is important in chemical tagging. In this case, the total mass in a survey sample coming from clusters within a mass bin $\delta M$, can be calculated to be 
\begin{equation}
M \, {\rm d}N/{\rm d}M \, \delta M = M^2 \, {\rm d}N/{\rm d}M \, \delta \log M \propto \delta \log M.
\end{equation}

\noindent
Quantitatively, this means that the chance of sampling a star from the logarithmic bin $[10 \, M_\odot,100 \, M_\odot]$ is the same as the probability of sampling from the logarithmic bin $[100 \, M_\odot, 1000 \, M_\odot]$, and so forth. Since we adopt a maximum cluster mass $M_{\rm cluster}^{\rm max} = 10^5 - 10^7 \, M_\odot$ in this model, we have $4-6$ orders of dynamical range in the cluster mass. This large range of cluster mass implies that clusters with $[10 \, M_\odot,100 \, M_\odot]$ contribute only $\sim 10\% - 25\%$ of the total stellar mass. \citet{lad03} determined that the CMF low mass cutoff occurs around $M_{\rm cluster} = 50 \, M_\odot$, which we will adopt as the fiducial value. 

Although not shown in this paper, we find that changing the low mass cutoff to $10 \, M_\odot$ or $100 \, M_\odot$ has a negligible effect on the results. First, as we have discussed, the small clusters only contribute $\sim 10\%-25\%$ of the population. Furthermore, changing the low mass cutoff will alter the number of small clusters and hence the background in each cell, however since the signal is concentrated in $\sim 0.1\%-1\%$ of the chemical cells, as we will show in \S\ref{subsec:local-results}, only $< 1\%$ of this background change is affecting the signal.

The high mass cutoff $M_{\rm cluster}^{\rm max}$ has a dramatic effect on the results because massive clusters dominate the signal, as shown in later sections. We therefore consider several different scenarios for the high mass cutoff and its evolution with redshift (see the lower right panel of Figure \ref{fig:model-properties}). The largest open clusters observed in the Milky Way appear to be Westerlund 1 \citep[e.g.,][]{bra08}, Berkeley 39 \citep[e.g.,][]{bra12} and Arches \citep[e.g.,][]{esp09}, with a mass few times of $10^4 \, M_\odot$. Noting the fact that the cluster could have gone through a period of rapid mass loss in its formation phase \citep[e.g.,][]{lad03}, we adopt $M_{\rm cluster}^{\rm max} \simeq 10^5 \, M_\odot$ at $z=0$ as the nominal mass cutoff at $z=0$ in the Milky Way disk.

A number of arguments suggest that the CMF high mass cutoff could have been higher in the past. For instance, the existence of massive globular clusters with surviving mass of $10^{4.5} - 10^{6.5} \, M_\odot$ \citep[e.g.,][]{har94} suggests that early conditions in the Galaxy favored the formation of more massive clusters.  Observations of high-redshift disk galaxies also suggests a high frequency, relative to $z=0$, of very massive gas clumps of $10^7-10^9 M_\odot$ \citep[e.g.,][]{gen06,for09,jon10,liv12}.

\citet{esc08} provided a simple model for the maximum cluster mass by studying gravitational instability in disks, similar to Toomre's classic analysis \citep{too64}. They calculate the maximum unstable mass to be $M_{\rm cluster}^{\rm max} = \Sigma_{\rm gas} (\lambda_{\rm rot}/2)^2$, where $\lambda_{\rm rot} = \pi^2 G \Sigma_{\rm gas}/\Omega^2$. From this formula, they further found that the maximum cluster mass can be determined by the gas fraction $\eta$ (i.e., gas mass to the total gravitational mass) and the total gas mass $M_{\rm gas}$ alone, where
\begin{equation}
M_{\rm cluster}^{\rm max} \propto M_{\rm gas} \eta^2.
\end{equation}

\noindent
The normalization of this formula depends on a variety of unknown parameters and so we choose instead to fix the normalization by hand at $z=0$. The dynamics of the Milky Way disk can be explained without appealing to dark matter, at least within the Solar radius. We therefore ignore the influences of dark matter when computing the upper mass cutoff, i.e., we define $\eta = M_{\rm gas}/(M_{\rm gas} + M_\star)$. The evolution of $M_{\rm gas}$ and $M_\star$ follow the discussion in \S\ref{subsec:sfh} and \ref{subsec:gas-mass}.

We consider three scenarios for the evolution of the upper mass cutoff, which we will denote as the quiescent, fiducial and optimistic models (see Figure \ref{fig:model-properties}). In the quiescent model, we consider the fiducial SFH and fix $M_{\rm cluster}^{\rm max} (z) = 10^5 \, M_\odot$ through cosmic time. In the fiducial and optimistic cases, we consider the SFHs labeled as fiducial and optimistic in Figure \ref{fig:model-properties} and allow $M_{\rm cluster}^{\rm max} (z)$ to evolve. We set $M_{\rm cluster}^{\rm max} (z=0) = 10^5 \, M_\odot$ for the fiducial case, and $M_{\rm cluster}^{\rm max} (z=0) = 3 \times 10^5 \, M_\odot$ for the optimistic case. We use the term ``optimistic'' because this model allows the formation of very massive clusters, which is favorable for chemical tagging. Finally, we impose a maximum upper limit of $10^7 \, M_\odot$. Clusters with mass larger than this cutoff are unlikely to be homogeneous \citep{bla10b} in their elemental abundances due to self-enrichment. The evolution of $M_{\rm cluster}^{\rm max} (z)$ in these three cases are plotted in the bottom right panel in Figure \ref{fig:model-properties}. The main differences of these three CMF models are summarized in Table \ref{table:models}. The range of CMFs we consider is similar to the range explored by \citet{bla10a}, although the authors do not consider a time-dependent CMF as we do here (for the optimistic and fiducial models).

%
%
%
%
%
%
\subsection{Chemical space}
\label{subsec:chem-model}

The last model ingredient is multi-dimensional space of elemental abundances, often referred to as the ``chemical space''. The chemical space is spanned by the elemental abundances [Fe/H], [$X_1$/Fe], $\ldots$, [$X_n$/Fe], where $X_1$ to $X_n$ are $n$ different elements measured. Since stars that were born together are expected to share the same abundances, they should reside at the same location in chemical space. 

As we will show below, the number of chemical cells in chemical space $N_{\rm cells}$ is a key variable in chemical tagging. To understand its importance, let's consider the case where we have an infinite number of chemical cells, in other words we have infinite resolution in the chemical space. In this case, all clusters from various birth sites can be easily identified. However, as the number of cells decreases, the probability that two clusters occupy the same chemical cell increases. In this case, the smaller clusters (in terms of the number of stars sampled per cluster) become contaminants in the detection. They dilute the number of genuine members of the massive clusters.

$N_{\rm cells}$ depends on two ingredients: (a) The chemical space spanned by the sample. This volume is governed by Galactic chemical evolution and survey design, including the number of elements of each star the survey can extract. Note that the volume does not scale in a simple way with the number of elements measured because of the strong correlation between various subgroups of elements. (b) The abundance measurement uncertainty $\sigma_{[X/{\rm Fe}]}$, which sets the volume of each cell. Regarding (b), in this study, we assume that the width of chemical cell is $1.5 \, \sigma$, i.e., two different distinct groups in chemical space can be recovered if their separation is larger than $1.5 \, \sigma$, where $\sigma$ represents the uncertainties along the principal components/independent dimensions.\footnote{As these component vectors are comprised of various elements, the uncertainties along these directions require the full covariance matrix of $\sigma_{[X/{\rm Fe}]}$.} Note that, given a chemical space of $N_{\rm dim}$ (independent) dimensions, the volume of each cell is proportional to $\sigma^{N_{\rm dim}}$. As a consequence, the number of cells is extremely sensitive to the abundance measurement uncertainties. We therefore stress that not only are small uncertainties favorable, but also accurate measurement of the uncertainties and their covariances are equally important.

The chemical space spanned by the sample, in principal, can be modeled through chemodynamical simulations. However, we note that chemical evolution models are still rather uncertain for many elements and are often limited to a relatively small number of elements \citep[e.g.,][]{kob06, min13}. \citet{kob11} include more elements, but they do not include neutron capture elements. Therefore, we are not aware of an existing chemical evolution model that encompasses all $\sim 25$ elements measured by the GALAH and Gaia-ESO surveys. For these reasons, and for simplicity, we choose here to adopt empirical results in estimating the volume and defer a chemical modeling approach to future work.

We make use of the estimated chemical space volume of Milky Way disk stars from \citet{tin12a} \citep[also see][for a similar study on bulge stars]{and12}. Using principal components analysis, \citet{tin12a} searched for directions in the chemical space that are orthogonal to each other and contain most variances of the data. These principal components define a n-dimensional cube spanned by the data. By definition, the number of cells is the volume of the cube divided by the volume spanned by each cell. As for the latter, given the assumption that the width of chemical cell is $1.5 \, \sigma$, the volume of the chemical cell is $(1.5 \, \sigma)^{N_{\rm dim}}$. The volume of the n-dimensional cube can be estimated from the width of edges in each dimension, which can be calculated from the principal components axial ratios.  Here we use the axial ratios of the principal components to estimate the volume that will be spanned by the GALAH data, as an example. The axial ratios of the first 6 dimensions are 1, 0.4, 0.25, 0.25, 0.1, 0.1. Apart from the obvious additional dimension from [Fe/H], \citet{tin12a} speculated that there should be another dimension associated with neutron capture elements. This last dimension was not available in the data analyzed by Ting et al. but will be probed by both GALAH and Gaia-ESO. 

We can safely assume that the first principal component spans at least 1.5 dex as it is the diagonal direction of the 17 dimension in study. Let's further assume that [Fe/H] and both of the additional dimensions span $1 \, {\rm dex}$, and the uncertainties along the independent dimensions are $\sigma = 0.1\,$dex. A simple calculation using the axial ratios yields: $N_{\rm cells} = (1.5 \, {\rm dex})^6 \times (1 \cdot 0.4 \cdot 0.25 \cdot 0.25 \cdot 0.1 \cdot 0.1) \times (1 \, {\rm dex})^2/(1.5 \, \sigma)^8= 10^4$ for GALAH. The Gaia-ESO survey spans a comparable list of elements and should therefore contain a similar number of $N_{\rm cells}$. An APOGEE-like survey should have $2-3$ fewer independent dimensions than GALAH \citep{tin12a}. All other parameters being the same, APOGEE should have $N_{\rm cells} \sim 10^3$. 

The above calculations are simple estimates for the number of chemical cells that could easily be off by an order of magnitude. Hence, in the analysis below we consider a wide range in this important parameter, ranging from $10^3-10^5$.
\begin{figure*}
\center
\includegraphics[width=\textwidth,natwidth=1400,natheight=1050]{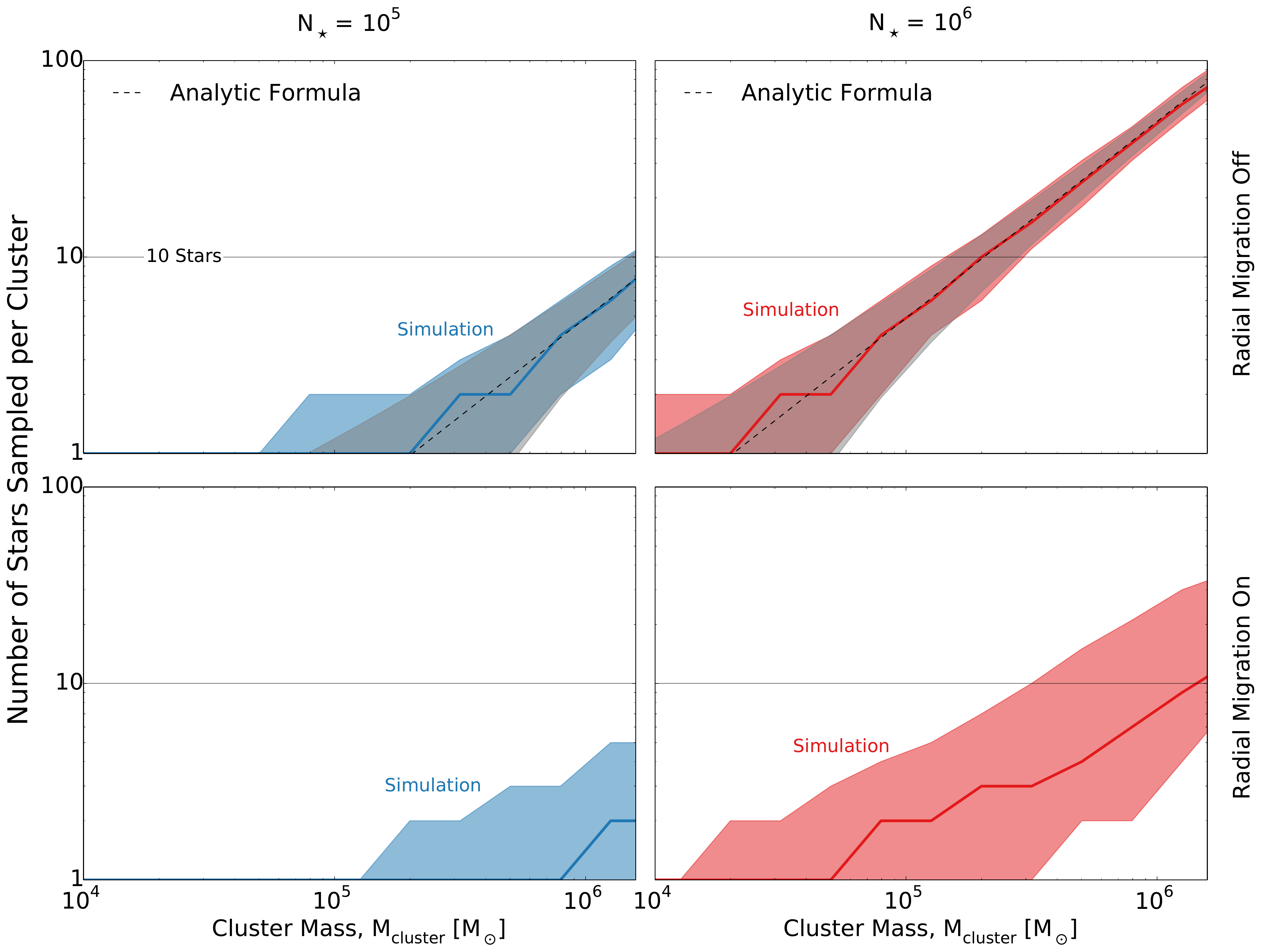}
\caption{Number of stars sampled per cluster as a function of cluster mass, assuming $\Delta R_{\rm survey} = \pm 3 \, {\rm kpc}$. The left panels assume $N_\star = 10^5$, whereas the right panels assume $N_\star = 10^6$. The top panels show the cases where there is no radial migration ($f_{\rm in-situ} = 100\%$), while the bottom panels illustrate the cases with radial migration and an in-situ fraction $f_{\rm in-situ} = 50\%$. The solid lines show the median and the shaded regions in color show the $1\sigma$ range of the results from simulations. In the limit of no radial migration, the number of stars sampled per cluster can be predicted analytically from equation (6). The predictions from the analytic formula are shown in dashed lines and gray shaded regions. The $1\sigma$ range from simulations follows very well the Poisson expectations. However, the analytic formula does not work in the case with radial migration because ex-situ clusters tend to have fewer stars sampled and bring down the number (see text and Figure \ref{fig:average-num-detail} for details).}
\label{fig:average-num}
\end{figure*}

%
%
%
%
%
%

\section{Results}
\label{sec:results}

With the model for the Milky Way disk stars now in hand, we turn to using that model to explore what ongoing and future massive spectroscopic surveys of stars may expect to reveal in the context of chemical tagging. In \S\ref{subsec:average-num} we investigate how many stars we expect to sample from the same cluster for different number of stars surveyed and both with and without the effect of radial migration. The main results are presented in \S\ref{subsec:local-results}, where we simulate the number of detectable groups in different scenarios. We study how observations of the distribution of stars in chemical space may encode information on the shape of the CMF. We also investigate whether each detectable group in chemical space is dominated by a single cluster or is comprised of a wide range of clusters.

%
%
%
%
%
%
\subsection{Number of stars sampled per cluster}
\label{subsec:average-num}

In this section we study the number of stars sampled per cluster for several idealized surveys. In particular, we are interested in how many stars will be sampled per cluster after the cluster is dispersed and mixed with the background sea of other clusters, and how the process of radial migration influences the sampling. Note that since we consider quantities as a function of cluster mass in this section, for a fixed $\Sigma_\star (R_0,z=0)$ the results will be independent of the CMF. However, the results do depend on $\Delta R_{\rm survey}$ and $f_{\rm in-situ}$ as these parameters change the sampling rate and the radial migration prescription. Here we assume $\Delta R_{\rm survey} = \pm 3 \, {\rm kpc}$ and $f_{\rm in-situ} = 50 \%$.

In Figure \ref{fig:average-num}, we plot the number of stars sampled per cluster as a function of cluster mass. The solid lines show the median of the results in each cluster mass bin and the shaded color regions show the $1\sigma$ range. In the top panels, we consider the case without radial migration, i.e., stars stay in the orbiting radii that they formed, while the bottom panels show the case with radial migration. The left and right panels show results for $N_\star = 10^5$ and $N_\star = 10^6$. A horizontal line at $N=10$ stars is meant to serve as a reference point.

While the results in Figure \ref{fig:average-num} clearly show that the {\it typical} sampling rate (within $\pm 1\sigma$ range) per cluster is quite low, except in the case of large $N_\star$ and high cluster mass, we emphasize that the {\it distribution} of the number of stars sampled per cluster has a long tail toward high values. We return to this point below.

\begin{figure}
\center
\includegraphics[width=0.45\textwidth]{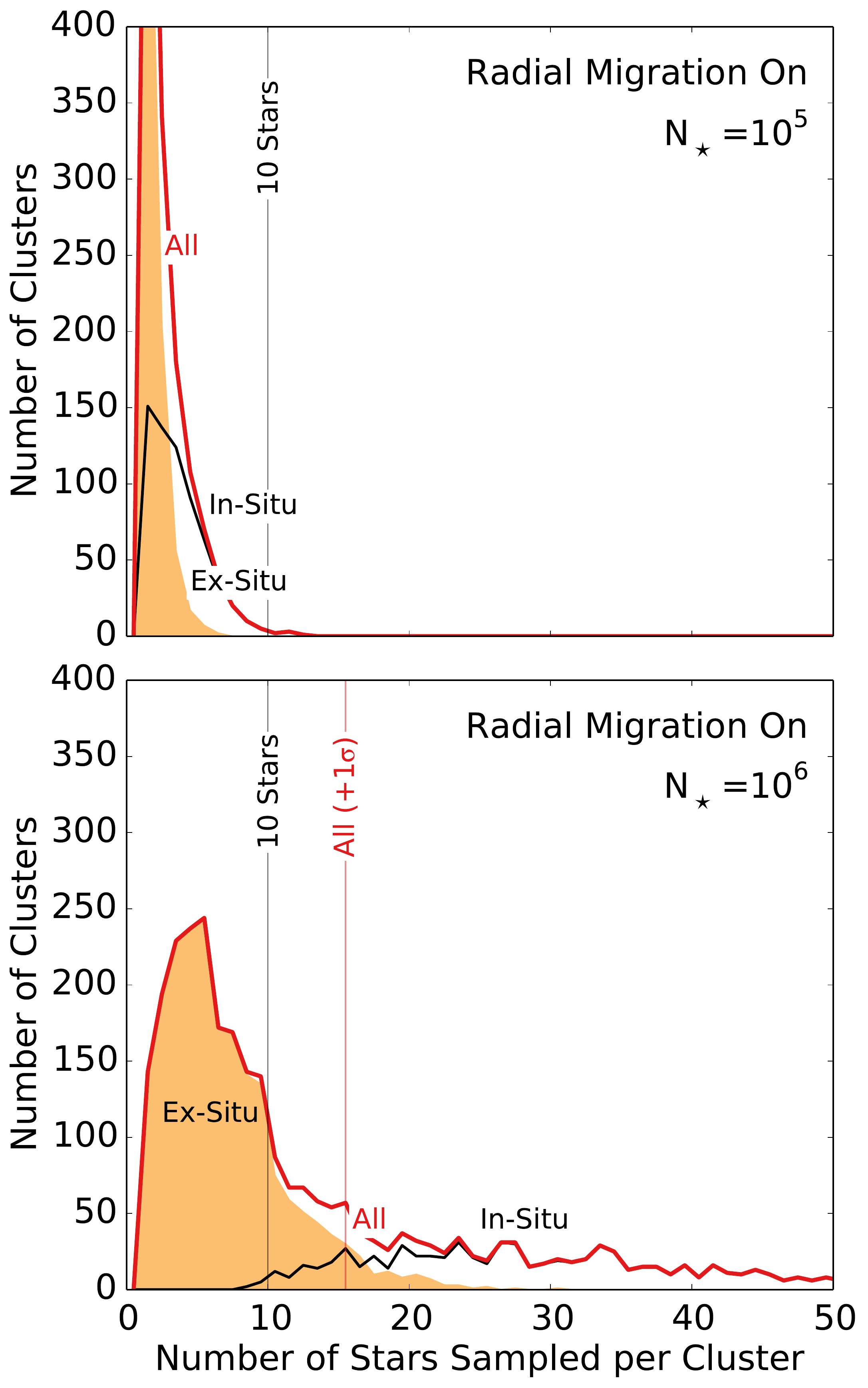}
\caption{Distribution of the number of stars sampled per cluster for $M_{\rm cluster} = (0.7 - 1.3) \times 10^6 \, M_\odot$. The top panel shows the result for $N_\star = 10^5$ and the bottom panel shows $N_\star = 10^6$. We assume $\Delta R_{\rm survey} = \pm 3 \, {\rm kpc}$ and $f_{\rm in-situ} = 50 \%$. We separate the cluster population into two - the in-situ and ex-situ populations. The ex-situ clusters have much smaller number of stars sampled per cluster compared to the in-situ population, indicating that ex-situ stars are mostly contaminants in chemical tagging. The red vertical line shows the 75 percentile of the combined results from in-situ and ex-situ clusters.}
\label{fig:average-num-detail}
\end{figure}

In the limit where there is no radial migration, the average number of stars (with $\langle M \rangle =1 \, M_\odot$) sampled per cluster can be analytically derived \citep[see also][]{des15}. The number of stars sampled per cluster is simply
\noindent
\begin{equation}
N_{\rm cluster} = M_{\rm cluster} \frac{N_\star}{M_{\rm annulus}}.
\end{equation}

\noindent
Recall that $M_{\rm annulus}$ is the total integrated SFR in the Solar annulus and $N_\star/M_{\rm annulus}$ is proportional to the sampling rate. This analytic model is shown in the top panels of Figure \ref{fig:average-num} and clearly predicts very well the results of the simulations. The grey shaded region demarks the $1\sigma$ from this analytic model.
\begin{figure*}
\center
\includegraphics[width=\textwidth,natwidth=1400,natheight=500]{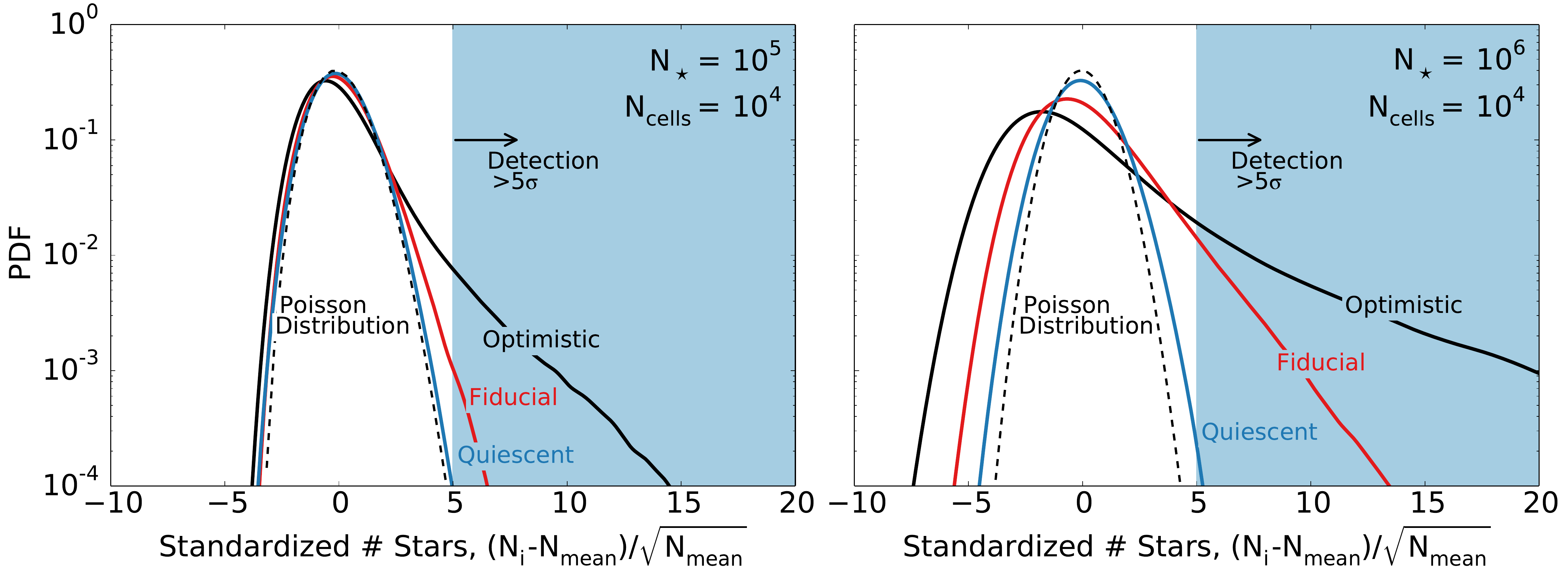}
\caption{Standardized number of stars in each cell compared to a Poisson distribution, where the mean of Poisson distribution is $N_{\rm mean} = N_\star/N_{\rm cells}$ and the standard deviation follows $\sigma = \sqrt{N_{\rm mean}}$. Cells in which the number of stars sampled exceeds $5\sigma$ are considered as detectable groups. The y-axis shows the probability of a detected group having a certain deviation from the Poisson distribution, quantified by the standardized number of stars. The integral under each curve is one. Unless stated of otherwise, we assume fiducial values for all the model parameters, as listed in Table~\ref{table:parameters}. Different CMFs show different degrees of deviation from Poisson statistics. The clumpiness of the chemical space may therefore be a useful tool to probe the underlying CMF. }
\label{fig:local-poisson}
\end{figure*}

Although illustrative, this analytic formula is unfortunately not applicable when radial migration is included. First, radial migration increases the number of stars that could end up in the Solar annulus, which has the effect of increasing the effective volume of the survey. We can define an effective radius of the observed annulus to be the mean distance, 
\begin{equation}
R_{\rm effective} = \frac{1}{n} \sum_{i=1}^n |R_{i,{\rm birth}} - R_0|,
\end{equation}

\noindent
where we sum over all the stars in the Solar annulus at the present-day. This equation takes into account the fact that, with radial migration, the actual sampled volume is larger than the observed volume because $|R_{i,{\rm birth}} - R_0| \geq |\Delta R_{\rm survey}|$. The effective integrated SFR $M_{\rm annulus}'$ within this effective volume is strictly larger than the one without radial migration due to the migration of ex-situ population, and therefore the number of stars per cluster will generally be lower than in the case without radial migration.

Moreover, clusters that were born ex-situ are unlikely to have a significant number of stars migrated into the Solar annulus. As shown in Figure \ref{fig:radial-migration}, while stars born $5 \, {\rm kpc}$ from the Galactic center can move into the Solar annulus at $R_0 = 8 \, {\rm kpc}$, only a small fraction of this population is in the Solar annulus. Figure \ref{fig:radial-migration} suggests that most of the ex-situ stars, even from massive clusters, will tend to enter as ``contaminants'' in the sense that they will have only $\mathcal{O}(1)$ stars sampled per cluster. In addition, some stars that were born in-situ will migrate outside the Solar annulus, further diluting the number of members of in-situ clusters. All of these effects work in the same direction of reducing the number of stars per cluster compared to a model without radial migration.

In Figure \ref{fig:average-num-detail} we show the distribution of the number of stars sampled per cluster for two choices of $N_\star$. This figure shows the distribution for a vertical slice in Figure \ref{fig:average-num} at a cluster mass of $\sim10^6\,M_\odot$. By separating the in-situ and ex-situ populations, Figure \ref{fig:average-num-detail} shows that the ex-situ population has on average a much smaller number of stars sampled per cluster, in agreement with the arguments described above. Although not shown, we checked that the in-situ population is only marginally influenced by radial migration --- only a small fraction of in-situ stars leave the Solar annulus. The mild effect on in-situ clusters is likely due to the fact that we consider a fairly large Solar annulus width of $\Delta R_{\rm survey} = \pm 3 \, {\rm kpc}$. In the radial migration prescription in this study, a typical radial migration length is $\sim 2 \, {\rm kpc}$, which is smaller than $|\Delta R_{\rm survey}|$. Although the typical radial migration length is still largely unconstrained from observations, some studies have suggested that since $R_0$ is beyond the outer Limblad resonance of the Galactic bar \citep{deh00}, a typical radial migration length is $< 2 \, {\rm kpc}$ \citep{hal15}.

Another feature evident in Figure \ref{fig:average-num-detail} is the tail of clusters with a large number of stars sampled per cluster. This highlights that median statistics are not sufficient to capture the full variety of expected behavior. These rare clusters may end up being the most valuable from the standpoint of chemical tagging as they should stand out as strong concentrations of stars in chemical space. The following section explores this effect in detail.

%
%
%
%
%
%
\subsection{Finding and counting groups in chemical space}
\label{subsec:local-results}

Observational uncertainties on elemental abundances impose a finite resolution in chemical space that can have important consequences for chemical tagging \citep{bla10a}. In this section, we simulate observational results by studying detections on a chemical cell-by-cell basis. In the following, for each generated sample, we distribute sampled clusters uniformly (on average) into $N_{\rm cells}$ cells. We perform Monte Carlo simulations and take the mean from 100 realizations. By jack-knife estimation, we find that the uncertainties on the mean is $\ll 10\%$ for $N_{\star} = 10^5 - 10^6$.

We define several terms that will be important in this section. A cell that contains a high density of stars compared to the mean defines a ``group''. We distinguish between ``group'' and ``cluster'' because the former can be comprised of multiple clusters. The cluster with the most stars sampled in each cell is referred to as the dominant cluster. Stars from the dominant cluster define the ``local signal''. The rest of the stars in the cell are referred to as ``local noise''.

%
%
%
%
%
%
\subsubsection{Identifying groups in chemical space}
\label{subsec:search-poisson}

If we were to randomly distribute $N_\star$ stars into $N_{\rm cells}$ chemical cells, the number of stars per cell should follow a Poisson distribution with a mean $N_{\rm mean} = N_\star/N_{\rm cells}$ and a $1\sigma$ range of $\sqrt{N_{\rm mean}}$. Since stars are born in clusters, there will be clumping in chemical space that is larger than Poisson expectations. The degree of clumpiness depends on several factors, chief among them is the form of the CMF \citep{bla10a}.

Operationally we define a cell as containing a ``detected'' group of stars if that cell deviates from Poisson expectations by at least $5\sigma$ and the total number of stars in that cell $>1$. Figure \ref{fig:local-poisson} shows the deviations from Poisson statistics for different CMFs and numbers of stars in the survey. In the right panel, we assume $N_\star = 10^6$. In this case, both the fiducial and optimistic CMFs show substantial numbers of cells exceeding $5\sigma$ from the average. By contrast, when $N_\star = 10^5$ (left panel), only the optimistic CMF shows substantial deviation from Poisson expectations. 

Figure \ref{fig:local-poisson} demonstrates that the deviation from Poisson is minimal for a quiescent CMF. This lack of deviation is not unexpected because clusters with $M_{\rm cluster} < 10^5 \, M_\odot$ have $\mathcal{O}(1)$ stars detected per cluster even for $N_\star = 10^6$ (see Figure \ref{fig:average-num}). Hence, randomly distributing clusters in $N_{\rm cells}$ cells for a quiescent CMF is close to randomly distributing $N_\star$ in $N_{\rm cells}$ cells. 

Figure~\ref{fig:local-poisson} also shows that the distribution of deviations can be a sensitive probe of the CMF. CMFs with a higher mass cutoff produce more clumpiness in chemical space. Although not shown, a flatter CMF also entails a larger number of massive clusters and hence a clumpier chemical space, echoing the results of \citet{bla10a,bla14}. The effect of the CMF on the distribution of deviations could potentially be exploited to reconstruct the CMF (and the physical processes that the CMF depends on, such as the SFH) from observational samples. This will be the subject of future work.

%
%
%
%
%
%
\subsubsection{What are groups in chemical space comprised of?}
\label{subsec:not-individual}

\begin{figure*}
\center
\includegraphics[width=\textwidth,natwidth=1400,natheight=500]{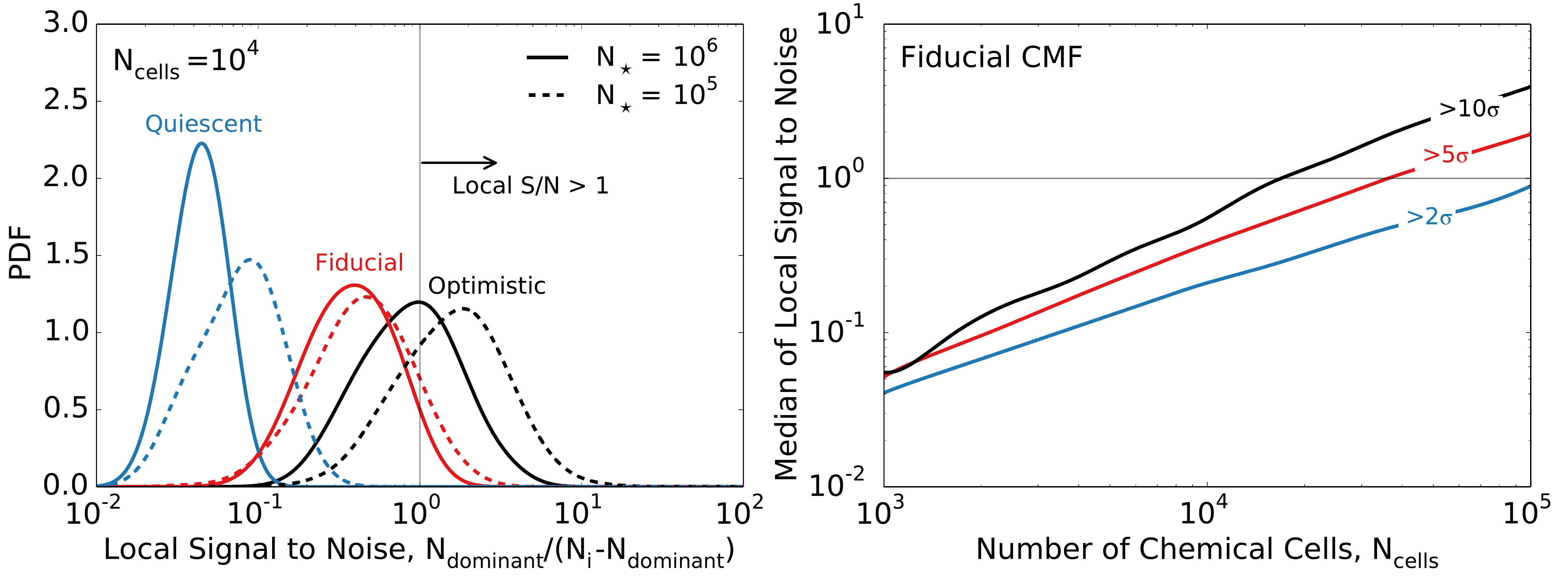}
\caption{{\em Left panel}: Local S/N ratio in chemical cells with $5\sigma$ more stars than the average. The number of stars sampled from the dominant cluster is considered signal in each cell, whereas the rest are considered noise. The y-axis shows the probability of a detected group having a certain local S/N. The integral under each curve is one. We assume $N_{\rm cells} = 10^4$. In this case, most detectable groups have local S/N $< 1$, showing that at least half of the stars in the detectable groups are not from dominant clusters. The difference between $N_\star = 10^5$ and $10^6$ is small, illustrating that sampling more stars increases the number of stars per cell, but it does not change the S/N. {\em Right panel:} Median of local S/N for different $N_{\rm cells}$. We assume a fiducial CMF in this panel. Unlike $N_\star$, increasing $N_{\rm cells}$ boosts the local S/N, and hence increases the chance of recovering individual clusters through chemical tagging.}
\label{fig:local-poisson-2}
\end{figure*}

In this section we investigate the properties of the``detected'' groups in chemical space (consisting of $>5 \sigma$ fluctuations). Figure \ref{fig:local-poisson-2} shows the distribution of the local ``S/N'' for those cells exceeding $5\sigma$ from Poisson statistics. Recall that the local S/N is defined as the ratio of stars coming from the most massive cluster in the cell to the remaining stars in that cell. A cell dominated by a single massive cluster will have high local S/N. In the left panel, we assume $N_{\rm cells} = 10^4$ and consider three different CMFs. Clearly most of the detected groups have local S/N $< 1$, especially for the quiescent and fiducial CMFs. 

This result is not surprising in light of the mean number of stars per cell ($100$ for $N_\star=10^6$ and $N_{\rm cells} = 10^4$).  In this regime, in order for the S/N to be $\gg1$, we would require that a single dominant cluster contribute $\gg100$ stars in a particular cell.  However, as shown in Figure \ref{fig:average-num}, the average number of stars sampled per cluster for the most massive clusters is $\sim100$ for $N_\star=10^6$.  The relatively low sampling rate, combined with the high average number of stars per cell, essentially guarantees that the local S/N will never be much larger than one.  As we discuss in \S\ref{sec:key-params}, the prospects for finding higher local S/N cells can be improved by searching in regions of chemical space in which the mean number of stars per cell is low.

The result in the left panel of \ref{fig:local-poisson-2} is fairly insensitive to $N_\star$. Increasing $N_\star$ increases both the number of stars sampled per cluster and the ``background'' comprised of stars from small clusters and hence the local S/N is left largely unchanged. In fact, the local S/N slightly decreases as we increase $N_\star$. This is not unexpected. As $N_\star$ decreases, it becomes more difficult to exceed the Poisson threshold. Therefore for smaller $N_\star$, the clumping of detected groups are mostly comprised of more massive clusters (e.g., $\sim 10^7 \, \rm M_\odot$), which implies a better local S/N. By contrast, for a larger $N_\star$, the clumping could either be due to a massive cluster or a few moderately massive clusters (e.g., $\sim 10^4 - 10^6 M_\odot$). While the S/N is somewhat negatively impacted by increasing $N_\star$, the total number of detected groups greatly increases with increasing $N_\star$, as shown in \S\ref{subsec:parameter-effect}.

The right panel of Figure \ref{fig:local-poisson-2} shows the median local S/N as a function of the number of chemical cells. Increasing $N_{\rm cells}$ results in a dramatic (almost linear) improvement in the local S/N. An increase in $N_{\rm cells}$ results in a decrease in the local background while keeping the signal unchanged. This panel also shows the effect of changing the definition of a ``detected'' group from $2\sigma$ to $10\sigma$.  Increasing the threshold has a modest effect on the local S/N but of course has a dramatic effect on the total number of resulting detected clusters. Although not shown, we have explored the effect of varying the slope of the CMF from $\alpha=-2.0$ to $-1.5$. This has only a modest effect on the trends shown in Figure \ref{fig:local-poisson-2}.

Note that the ($5\sigma$) deviation with respect to Poisson statistics is measurable in reality as it only requires the expected average number of stars in each cell. On the other hand, the local S/N is not measurable.\footnote{For readers who want to understand the number of groups that consist mainly a dominant cluster (e.g., having local S/N $\geq 1$), we urge readers to explore the interactive online applet (see Appendix~\ref{sec:interactive} for details). In the applet, we allow users to impose a local S/N criteria.} In this paper we only define ``detected groups'' according to a measurable parameter, and we emphasize again that we use the term ``group'' rather than ``cluster'' when describing clumps in chemical space because of the effect discussed in this section. The ambiguity that can arise, even when a cell deviates by more than $5\sigma$ argues strongly that interpretation of the data from ongoing and upcoming surveys will require models such as the one presented in this work.

%
%
%
%
%
%
\subsubsection{Number of detectable groups as a function of model parameters}
\label{subsec:parameter-effect}

\begin{figure*}
\center
\includegraphics[width=\textwidth,natwidth=1400,natheight=1000]{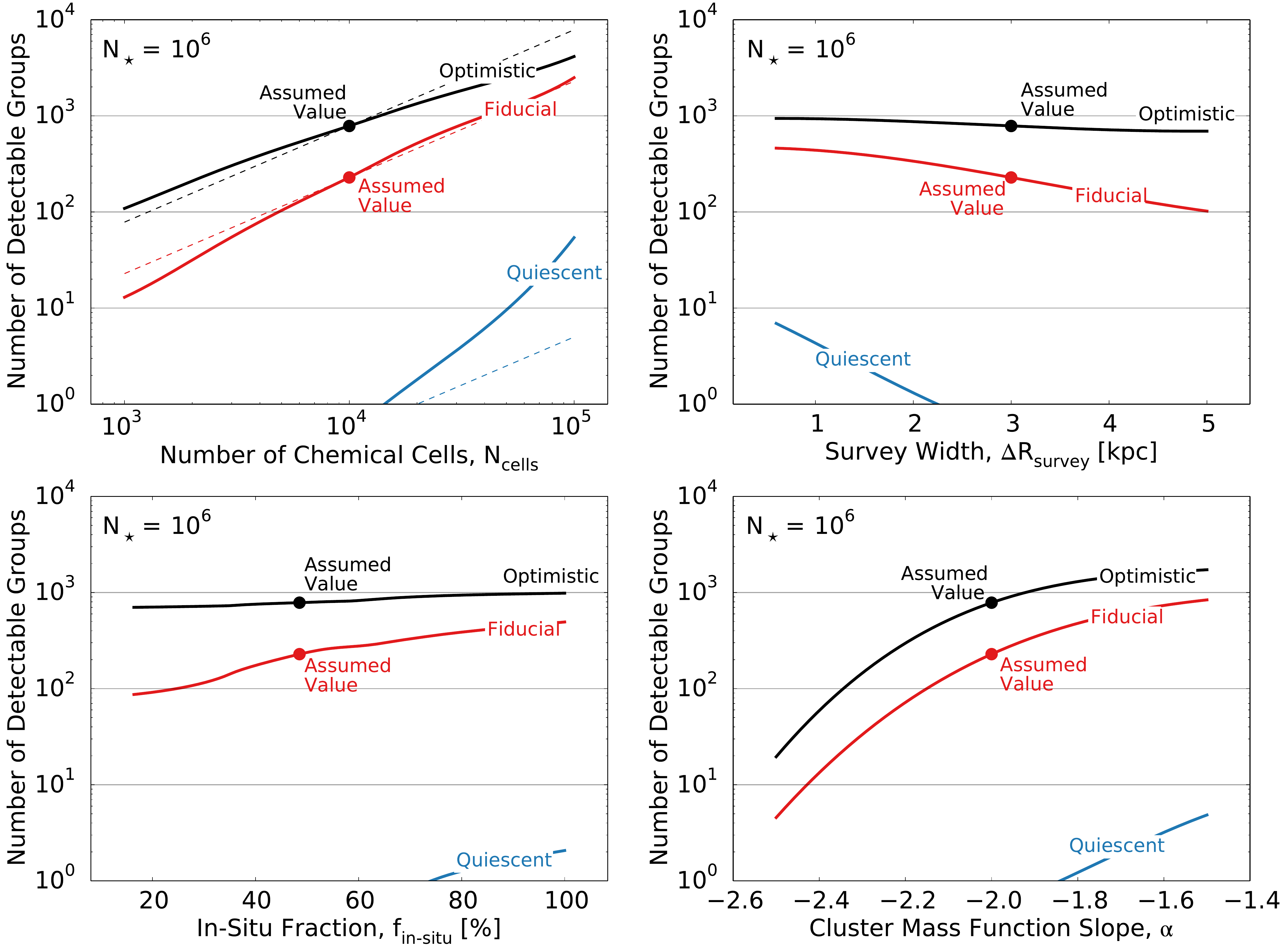}
\caption{Total number of cells that exceed $5\sigma$ from Poisson statistics as a function of a variety of model parameters. We vary each of these model parameters while fixing the rest to the fiducial values as listed in Table~\ref{table:parameters}. The three different solid lines show results from three CMF evolutions as illustrated in Figure \ref{fig:model-properties}. The dashed lines show linear relations for reference. The solid symbols show the results assuming fiducial values for all model parameters. See text for discussion.}
\label{fig:local-results-2}
\end{figure*}

\begin{figure*}
\center
\includegraphics[width=\textwidth,natwidth=1400,natheight=500]{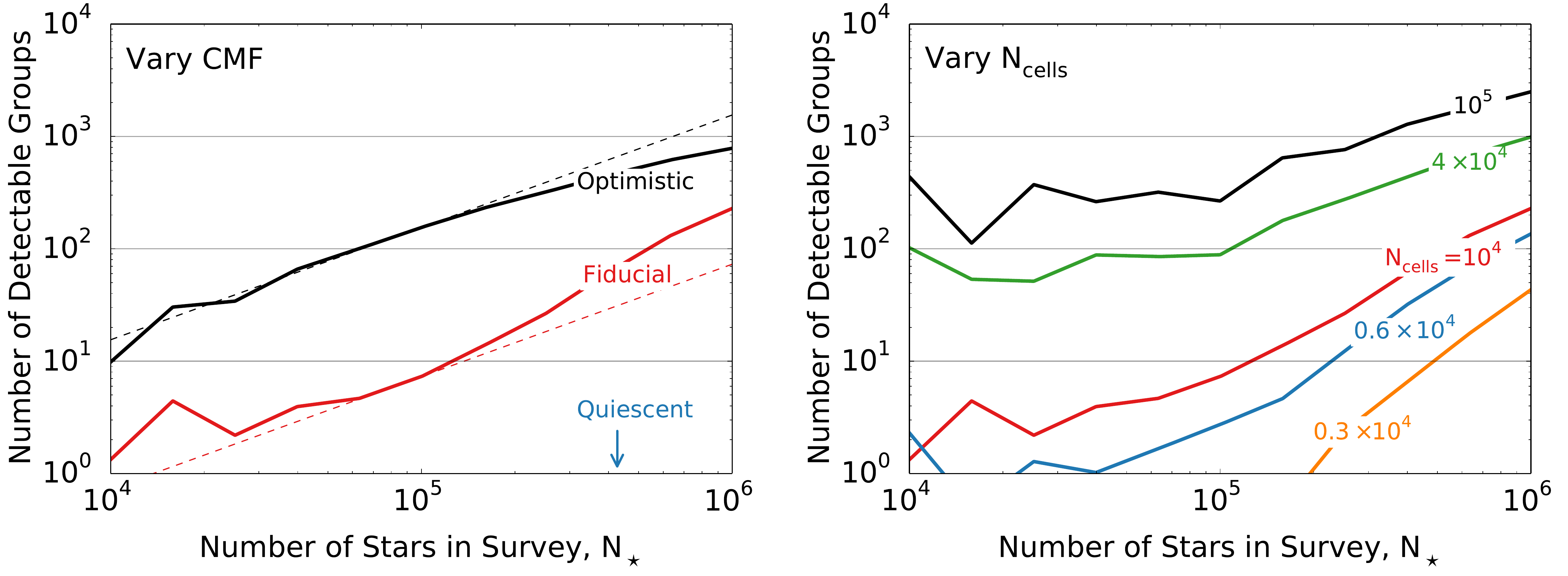}
\caption{Total number of cells exceeding $5\sigma$ from Poisson statistics as a function of the number of stars in the survey. We assume a survey width of $\Delta R_{\rm survey} = \pm 3 \, {\rm kpc}$ and $f_{\rm in-situ} = 50\%$. The red solid lines in both panels represent the reference results assuming a fiducial CMF and $N_{\rm cells} = 10^4$. The dashed lines show linear relations for reference. Different solid lines in the left panel show the results assuming different CMFs, whereas the right panel shows the results for different $N_{\rm cells}$. See text for discussion.}
\label{fig:local-results-3}
\end{figure*}

In this section we present the total number of detected groups in chemical space as a function of a variety of model parameters, including the in-situ fraction, $f_{\rm in-situ}$, CMF slope, $\alpha$, survey width $\Delta R_{\rm survey}$ number of chemical cells $N_{\rm cells}$, and number of stars in the survey, $N_\star$. We vary one of these model parameters at a time while adopting the fiducial values for the other model parameters (see Table~\ref{table:parameters}); modifying more than one parameters at once is allowed in the online applet. The results are presented in Figures \ref{fig:local-results-2} and \ref{fig:local-results-3}.

%
%
%
%
%
%
\noindent
\paragraph{Number of chemical cells}

As the number of chemical cells increases, more moderately massive (e.g., $\sim 10^4 - 10^6 M_\odot$) clusters start to occupy different cells instead of sharing the same cell. The total number of detectable groups thus increases, approximately linearly for the fiducial and optimistic CMFs. However, the gain is more drastic for CMFs with a smaller high mass cutoff. This trend is due to the fact that, given the same $N_\star$, moderately massive clusters are more abundant for CMFs with a smaller high mass cutoff. These clusters might not be detected with a smaller $N_{\rm cells}$. Including more cells benefits these moderate clusters the most. 

Since both the number of detectable groups and the local S/N (see \S\ref{subsec:not-individual}) are sensitive to $N_{\rm cells}$, it is clear that $N_{\rm cells}$ is one of the most important parameters in the context of chemical tagging. Recall that the number of cells scales as $\sigma^{-N_{\rm dim}}$, where $N_{\rm dim} \sim 8$ is the number of independent dimensions in the chemical space we can expect for upcoming optical surveys (GALAH and Gaia-ESO). Therefore, if we improve the abundance measurement uncertainties by a factor two, the number of chemical cells is improved by a factor of $2^8 \sim 250$. On the other hand, this also means that the number of chemical cells decreases by a factor $\sim2$ for every $10\%$ increase in the measurement uncertainties. Substantial effort should therefore go into decreasing (and characterizing!) the uncertainties in abundance measurements in upcoming spectroscopic surveys.

%
%
%
%
%
%
\noindent
\paragraph{Survey width}

As $\Delta R_{\rm survey}$ increases the number of detectable groups decreases. To understand this trend, it suffices to note that as we increase $\Delta R_{\rm survey}$ there are more stars in the annulus. As a result, the chance that we sample from the same cluster decreases (i.e., the sampling rate decreases). Since each cluster is sampled with fewer stars, the chance to observe signal spikes in chemical space also decreases. Therefore, the total number of detectable groups decreases as the survey width widens. In fact, since the volume of the Solar annulus is proportional to $\Delta R_{\rm survey}$, the number of stars in the annulus is also roughly proportional to $\Delta R_{\rm survey}$. Therefore, the sampling rate is, to first order, inversely proportional to $\Delta R_{\rm survey}$.

Interestingly, the survey width has less effect on CMFs with a larger higher mass cutoff. This trend is due to the fact that as we increase the survey width, we also increase the number of clusters, roughly in proportion to $\Delta R_{\rm survey}$. The most massive clusters are the least susceptible to change in sampling rate because a large number of stars from such clusters are already sampled in the fiducial case. For CMFs with a larger high mass cutoff, the decrease in sampling rate caused by an increase in $\Delta R_{\rm survey}$ is partly compensated by the increase in the number of massive clusters, resulting in a weak dependence of the number of detected groups on $\Delta R_{\rm survey}$.

%
%
%
%
%
%
\noindent
\paragraph{In-situ fraction}

As the in-situ fraction decreases, the number of cells exceeding $5\sigma$ decreases because there are more contaminants from ex-situ clusters (see Figure \ref{fig:average-num-detail}). However, the effect of in-situ fraction is rather marginal for CMFs with a larger high mass cutoff. This effect is best understood from Figure \ref{fig:local-poisson}. Most of the detectable groups for a quiescent CMF or a fiducial CMF are at the edge of the detection level of $5 \sigma$. Hence adding in additional background noise in the form of ex-situ stars can have a much larger effect for model with a quiescent CMF compared to an optimistic CMF, in which many of the cells far exceed the $5\sigma$ detection threshold.

%
%
%
%
%
%
\noindent
\paragraph{CMF slope}

As we vary the CMF slope, we are essentially redistributing mass between smaller clusters and massive clusters. This has two effects that act in tandem: a shallower CMF results in more massive clusters, which will have more stars sampled per cluster. In addition, a shallower CMF results in fewer low mass clusters that contribute primarily to the ``noise'' in a cell. The chemical space becomes much clumpier as $\alpha$ increases
\citep[also see][]{bla10a}, and as a result there are many more detected groups.

%
%
%
%
%
%
\noindent
\paragraph{Number of stars in the survey}

Since the number of stars sampled for massive clusters is roughly proportional to $N_\star$ while the Poisson threshold only grows as $\sqrt{N_{\rm mean}} \propto \sqrt{N_\star}$, increasing $N_\star$ improves the number of detectable groups, as shown in Figure \ref{fig:local-results-3}. In the left panel, the gain is approximately linear in $N_\star$ for the optimistic and fiducial CMFs. The right panel shows the gain in the number of detected groups as a function of $N_\star$ and $N_{\rm cells}$. The stochasticity at $N_\star \sim 10^4$ is likely due to the uncertainties in our Monte Carlo procedures.

%
%
%
%
%
%

\subsubsection{Selecting subpopulations}
\label{subsec:subpops}

As we argued in \S\ref{sec:overview}, the sampling rate, which is proportional to the number of stars in the survey divided by the number of stars in the survey volume, is a key parameter determining the number of stars sampled per cluster. In the limit where the sampling rate is 100\%, the main limiting factor for chemical tagging is the resolution in chemical space. One way to increase the sampling rate is to increase $N_\star$; this was discussed in the previous section. A second way is to decrease the number of stars in the survey volume. The latter will be effective only if one is able to identify a subpopulation of stars that corresponds to a subpopulation of clusters. For example, selecting on stellar age satisfies this criterion, while selecting a random subsample does not.

\begin{figure*}
\center
\includegraphics[width=\textwidth,natwidth=1400,natheight=500]{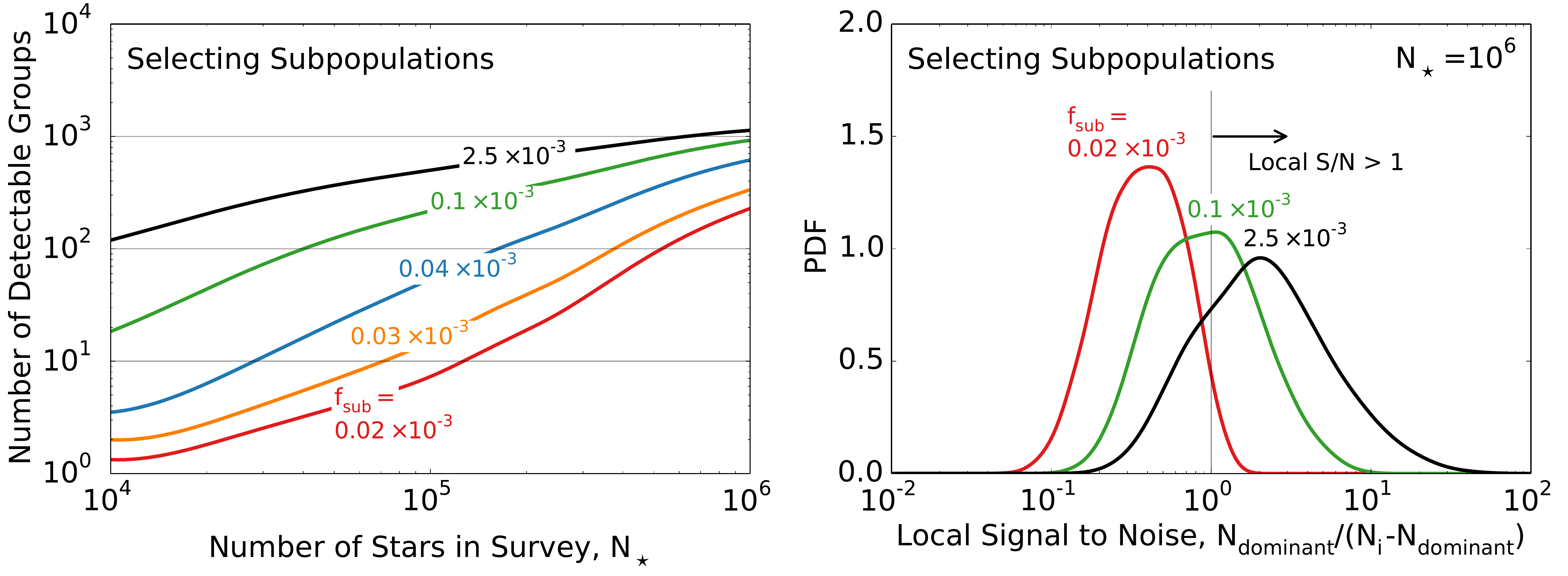}
\caption{{\em Left panel:} Total number of cells exceeding $5\sigma$ from Poisson statistics as a function of the number of stars in the survey. We assume a fiducial CMF, with $N_{\rm cells} = 10^4$, $\Delta R_{\rm survey} = \pm 3 \, {\rm kpc}$ and $f_{\rm in-situ} = 50 \%$. Different lines in this panel show the results assuming a variety of subpopulation selections. The subpopulations are selected through the stellar age criteria of $> 0 \, {\rm Gyr}$ (the lowest line), $> 3 \, {\rm Gyr}$, $> 6 \, {\rm Gyr}$, $> 9 \, {\rm Gyr}$ and $> 12 \, {\rm Gyr}$ (the highest line), respectively. The corresponding sampling rates, $f_{\rm sub}$, for $N_\star = 10^6$ are stated in each line. {\em Right panel:} Local S/N in each of the detected cells for different subpopulations, assuming $N_\star = 10^6$ and a fiducial CMF. The number of stars sampled from the dominant cluster is considered signal in each cell, whereas the rest are considered noise. See text for discussion.}
\label{fig:local-results-4}
\end{figure*}

Figure \ref{fig:local-results-4} considers the case where only stars above certain stellar ages are targeted in a survey. Since the number of older stars is smaller, there are not as many survey candidates compared to the case where we sample all disk stars uniformly. As a consequence, given the same $N_\star$, the chance that we sample from the same cluster improves. In addition to improving the total number of detectable groups, as we consider a more selective stellar subpopulation the number of clusters is reduced. The dominant cluster therefore contributes a greater fraction of the total stars in each detectable group because there are not as many clusters sharing the same cell. As shown in the right panel of Figure \ref{fig:local-results-4}, if the survey sample is collected randomly from all populations (the red solid line), most of the detectable groups have a local S/N of $0.3$. This local S/N value implies that only $0.3/(0.3+1) \simeq 25\%$ of the members of detectable groups are from the dominant cluster. However, if we only target old stars with stellar age $> 12 \, {\rm Gyr}$, the local S/N is $\sim 2$, indicating that $2/(2+1) \simeq 70\%$ members of each of the detectable groups are from the dominant cluster.

As a caveat, we caution that the interpretation of Figure \ref{fig:local-results-4} is complicated by the fact that the selection of older clusters also preferentially selects a population of stars forming from a CMF with a higher mass cutoff (at least for the fiducial model used in the figure). So not only is the sampling rate increasing but so also is the characteristic cluster mass. Future work is required to disentangle these effects.

%
%
%
%
%
%

\section{Discussion}
\label{sec:discussion}

%
%
%
%
%
%

\subsection{Summary of the key parameters affecting chemical tagging}
\label{sec:key-params}

The key parameters governing both the ability to detect groups in chemical space and the ``purity'' of those recovered groups (i.e., the local S/N) are the number of stars in the survey, $N_\star$, the number of chemical cells, $N_{\rm cells}$, the CMF, and the sampling rate. Table \ref{table:effects} presents a summary of the key variables and their effect on various quantities of interest.

Several of these parameters are either outside of the control of the observer, including the form and evolution of the CMF, or are trivially in control of the observer, such as $N_\star$. Others require further consideration. For example, the number of chemical cells depends on both the volume of chemical space and the size of each cell. The former depends on chemical evolution of the stellar population(s) under consideration, and can be influenced by the survey strategy. The latter is proportional to $\sigma^{-N_{\rm dim}}$ where $\sigma$ is the observational uncertainty on abundance measurements and $N_{\rm dim}$ is the number of effective dimensions in the chemical volume.

Perhaps the most conceptually complex parameter is the sampling rate. For a fixed $N_\star$ the sampling rate is inversely proportional to the total number of stars available within the survey design. The phrase ``survey design'' was chosen to highlight not only the survey volume but also the subpopulation under consideration. Moreover, with regards to the survey volume, this must be considered in an orbit-averaged sense. For example, a survey targeting stars within 1 kpc of the Sun has a survey volume in this definition that encompasses the entire annulus of the Galactic disk with a width of $\pm1$ kpc. Likewise, a pencil beam survey of bulge stars has a survey volume of the entire bulge. As we showed in \S\ref{subsec:subpops}, selecting subpopulations of stars can be very effective provided that the selection picks out a subset of clusters. Selecting on stellar age can achieve this, and so will effectively boost the average number of stars sampled per cluster. On top of that, selecting subsample reduces the number of clusters in each cell, and thus improves the local S/N in each detectable group. In contrast, a random subsample of stars will simply result in a smaller number of stars per cluster.

These parameters affect different aspects of chemical tagging. As shown in Table~\ref{table:effects}, increasing the number of stars or reducing the survey volume increases the number of detected groups and improves the reconstruction of the CMF because it increases the sampling rate, but it has little effect on the local S/N ratio. Even though the sampling rate increases in these cases, both the local signal and noise increase in similar proportions. In contrast, decreasing $\sigma_{[X/{\rm Fe}]}$ and/or selecting subpopulation reduces the average number of stars per cell, while maintaining the same signal. Therefore the local S/N improves as well.

In this work we focused on idealized surveys of stars in the Milky Way disk. In such situations the ratio of the number of stars in the annulus, $N_{\rm annulus}$ to $N_{\rm cells}$ is $\gg1$. However, there are regimes in which this ratio can be closer to or less than unity. \citet{bla10a} considered the regime of metal poor stars in dwarf galaxies. Such subpopulations could easily have a total number less than $N_{\rm cells}$. In this case the mean number of stars per cell will be $\ll1$ and so significant overdensities in chemical space will much more likely reflect a single cluster, rather than a superposition of multiple clusters \citep[see example in][]{kar12}. As argued by \citet{bla10a}, in this regime one can in principle find clusters in chemical space with a relatively modest number of stars surveyed, provided that the CMF is not too steep. Similarly, for a survey targeting disk stars, one might imagine the first chemical-tagging detections to come from the less populated regime in chemical space with a smaller contaminated background $N_{\rm mean}$ (i.e., outliers), as discussed in \citet{bla15}.

\begin{table}
\begin{center}
\caption{The effects of various survey strategies on chemical tagging detections.\label{table:effects}}
\begin{tabular}{lccc}
\tableline \tableline
\\[-0.2cm]
                                  & Improve the        & Improve chance  & Improve \\
                                  & number of          & of recovering   & reconstruction \\
                                  & detectable groups  & single cluster  & of CMF \\[0.1cm]
\tableline
\\[-0.2cm]
Increase $N_\star$                & \checkmark         &                 & \checkmark \\[0.1cm]
Decrease $\sigma_{[X/{\rm Fe}]}$  & \checkmark         & \checkmark      & \checkmark \\[0.1cm]
Reduce $\Delta R_{\rm survey}$    & \checkmark         &                 & \checkmark \\[0.1cm]
Subpopulations                    & \checkmark         & \checkmark      & \checkmark \\[0.1cm]
\tableline
\end{tabular}
\end{center}
\end{table}

%
%
%
%
%
%

\subsection{Strategies for optimizing the potential for chemical tagging}
\label{sec:design-survey}

The influence of key parameters on various observables allows us to consider ways in which one could optimize a spectroscopic survey of stars for the purposes of chemical tagging.

A survey that could reach $N_\star \sim 10^6$ and $N_{\rm cells} \gtrsim 4 \times 10^4$ could potentially achieve three major goals: (a) producing a sizable number ($\sim 10^3$) of detectable groups; (b) the detected groups would consist primarily of a single dominant cluster; and (c) reconstructing the CMF for $M_{\rm cluster}^{\rm max} \simeq 10^5 \, M_\odot$. These goals could be realized if the CMF is somewhere in the range between our ``fiducial'' and ``optimistic'' scenarios.
The GALAH survey \citep{des15} aims to observe $N_\star=10^6$; a key question will be whether or not the number of chemical cells is closer to $10^4$ or $10^5$ (see Section~\ref{sec:key-params} for the key dependencies).

Even if not all three goals are realized in the context of a massive spectroscopic survey, one could imagine a tiered approach. A survey of $10^6$ could be used to identify overdensities in the chemical space. One could then follow up those overdensities with higher quality spectroscopy to obtain more precise abundance constraints, or one could appeal to differential techniques to increase the relative abundance precision. One could also use other information to separate multiple clusters within a single cell, e.g., kinematics or color-magnitude diagrams.

Given that both $N_\star$ and $N_{\rm cells}$ affect the number of detected groups in chemical space in similar ways, is there an advantage to spending more time collecting greater numbers of stars, or more time obtaining higher quality spectra could lead to smaller $\sigma_{[X/{\rm Fe}]}$, more elements, and hence larger $N_{\rm cells}$? In the simplest scenario (assuming for example that one has not already exhausted the input catalog at a particular apparent magnitude), $N_\star$ is roughly proportional to the integration time. On the other hand, since $N_{\rm cells} \propto \sigma^{-N_{\rm dim}}$, there is an enormous gain in $N_{\rm cells}$ for even a modest improvement in the abundance uncertainties. For $N_{\rm dim} \sim 8$ independent dimensions (likely appropriate for e.g., GALAH), one could improve $N_{\rm cells}$ by a factor of two for a 10\% reduction in the abundance uncertainties (\S\ref{subsec:parameter-effect}). Therefore, if the goal is to find as many local peaks in chemical space (i.e., detectable groups) as possible and/or to increase the odds of those peaks being dominated by a single massive cluster, it might be more advantageous to seek strategies that reduce the abundance uncertainties rather than simply acquiring more stars.

An effective way to improve chemical tagging detections is by targeting a stellar subpopulation exclusively. As we have shown in \S\ref{subsec:subpops} and discussed in \S\ref{sec:key-params}, targeting a subpopulation not only improves the sampling rate but also reduces the number of clusters per chemical cell. It improves chances of the reconstructing the CMF because there are more stars sampled per cluster and more significant deviations from Poisson statistics. It also improves the local S/N and hence the chance of recovering individual clusters within detected groups in chemical space. 

A variety of properties could be used to select special subpopulations from a larger parent sample, including age, metallicity, and kinematics. One could envision pilot surveys at modest spectral resolution designed to select stars in a narrow range in [Fe/H]. Kinematics from Gaia could be used to separate hot and cold components, for example thin and thick disk stars \citep[e.g.,][]{red06}. Stars could also be selected according to their age once age measurements are available for large samples of stars, e.g., from isochrone fitting and/or asteroseismic constraints. Finally, in an optically selected survey such as GALAH, which is biased to higher Galactic latitudes, it preferentially observes thick disk stars \citep{des15}. Since the total number of thick disk stars is smaller than thin disk stars, this preference argues that the sampling rate in these surveys could be larger than the one we assume in this study as we adopt an uniform sampling strategy \citep[see also][]{bla15}.

%
%
%
%
%
%

\subsection{Caveats, limitations, \& future directions}

A variety of assumptions and simplifications were made in this study. Here we highlight the most important limitations and comment on future directions.

When populating the chemical space we assumed that clusters are (statistically) homogeneously distributed in all $N_{\rm cells}$ chemical cells available. From both observations and chemical evolution models we know that this assumption is not true in detail. Of course, there are many more high metallicity stars than low metallicity stars, but also we expect the size of the chemical space to vary systematically with metallicity (for example, due to certain nucleosynthetic pathways, e.g., in AGB stars, that only become important some time after the initial burst of star formation). Because of these complexities, the space cannot be completely described by the parameter $N_{\rm cells}$. A more accurate approach would be to include a model for chemical evolution and then to define overdensities in chemical space with respect to a local background, either using neighboring cells or a more sophisticated group finding algorithm \citep[e.g.,][]{sha09,mit13}. 

This study focused on idealized surveys targeting Milky Way disk stars. We did not consider the bulge, stellar halo, disrupted satellite galaxies, nor nearby dwarf galaxies. Each of these populations offers a unique set of challenges and opportunities. These components will be included in future versions of the model.

We did not follow the actual orbits of stars in a live Galactic potential, and the treatment of radial migration is quite simplistic. One could imagine an extension to the current model that follows the dynamical disruption of star clusters and the sequent orbital histories of the individual stars. This would be very valuable for exploring the potential gains of folding in kinematic information, such as will soon be available from Gaia and/or from the spectroscopic surveys themselves. \citet{mit14} found that kinematics information does not improve the detectability, but it is likely due to the limitation of their small sample with $< 10^3$ stars. As we have demonstrated in this study, detected groups in small sample are not likely to be co-natal, agreeing with their assessment.

The adopted model for the gas mass is fairly simplistic. However, we emphasize that the gas mass distribution only influences the radial migration prescription and the evolution of the CMF. The former is parameterized via the in-situ fraction, $f_{\rm in-situ}$. In both cases we consider a range of possible scenarios, which in some sense is equivalent to exploring the effects of varying the underlying gas mass model directly.

We assume that the spatial frequency of star formation follows an exponential disk characterized by the scale length $R_{\rm SFR}$. We are aware that this assumption might not be true in detail. At a given time, stars might form in some large scale molecular rings \citep[e.g.,][]{blo06,gor06} or spiral arms \citep[e.g.,][]{rix93,bik03}. However, we are only interested in the integrated star formation rate over the cosmic history. Since these transient complexes, at least for the molecular rings, are expected to be short lived and rapidly dissipate \citep[$< 100$ Myr; e.g.,][]{bas05,gor06}, the smooth star forming assumption is likely to do fine.

%
%
%
%
%
%

\section{Conclusions}
\label{sec:conclusions}

In this study we explored the prospects for chemically tagging stars in idealized spectroscopic surveys of the Solar vicinity. We constructed a simple two dimensional time-dependent model of the Milky Way disk including the effects of radial migration and evolution in the CMF. We explored a number of important parameters affecting the detectability of groups of stars in chemical space and we studied the composition of the detected groups. We now summarize our principle conclusions.

\begin{itemize}

\item The key parameters affecting the number of detected groups in chemical space, and whether or not those groups are dominated by a single massive cluster, are: the shape and evolution of the CMF; the number of chemical cells; and the survey sampling rate. The sampling rate is proportional to the number of stars in the survey divided by the total number of stars belonging to a particular (sub)population. The latter two parameters are strongly influenced by observational survey design choices.

\item The clumpiness in chemical space is strongly influenced by the CMF and by the survey sampling rate. This implies that one can probe the CMF of long disrupted clusters by statistically analyzing the clumpiness in chemical space.

\item Confidently identifying {\it individual} clusters through chemical tagging will be challenging even for $N_\star = 10^6$, if disk stars are uniformly sampled. Fundamentally this is because the sampling rate is inherently small in such cases ($\sim 10^{-4}$) implying that one expects to collect on average 10 stars per cluster for clusters with $M_{\rm cluster} \gtrsim 10^5 \, M_\odot$. This is born out by our modeling, where we find that even very large overdensities in chemical space are typically not comprised of stars from a single dominant cluster. In the fiducial case with $N_{\rm cells}=10^4$, the dominant cluster contributes only 25\% of the stars in the detected group. Additional follow-up of the stars within large overdensities in chemical space may provide additional discriminating power, either by decreasing the measurement uncertainties on the abundances, or by folding in color magnitude diagram or kinematic information.

\end{itemize}

%
%
%
%
%
%

\acknowledgments

The authors thank Joss Bland-Hawthorn, Ken Freeman, Sanjib Sharma, Charlie Lada and Eve Ostriker for helpful discussions, and Andrea Schruba and Peter Behroozi for sharing their data in electronic format. The computations in this paper were run on the Odyssey cluster supported by the FAS Division of Science, Research Computing Group at Harvard University.

%
%
%
%
%
%

\appendix

\section{Sampling Algorithm and Computational Cost}

\begin{figure}
\center
\includegraphics[width=0.45\textwidth]{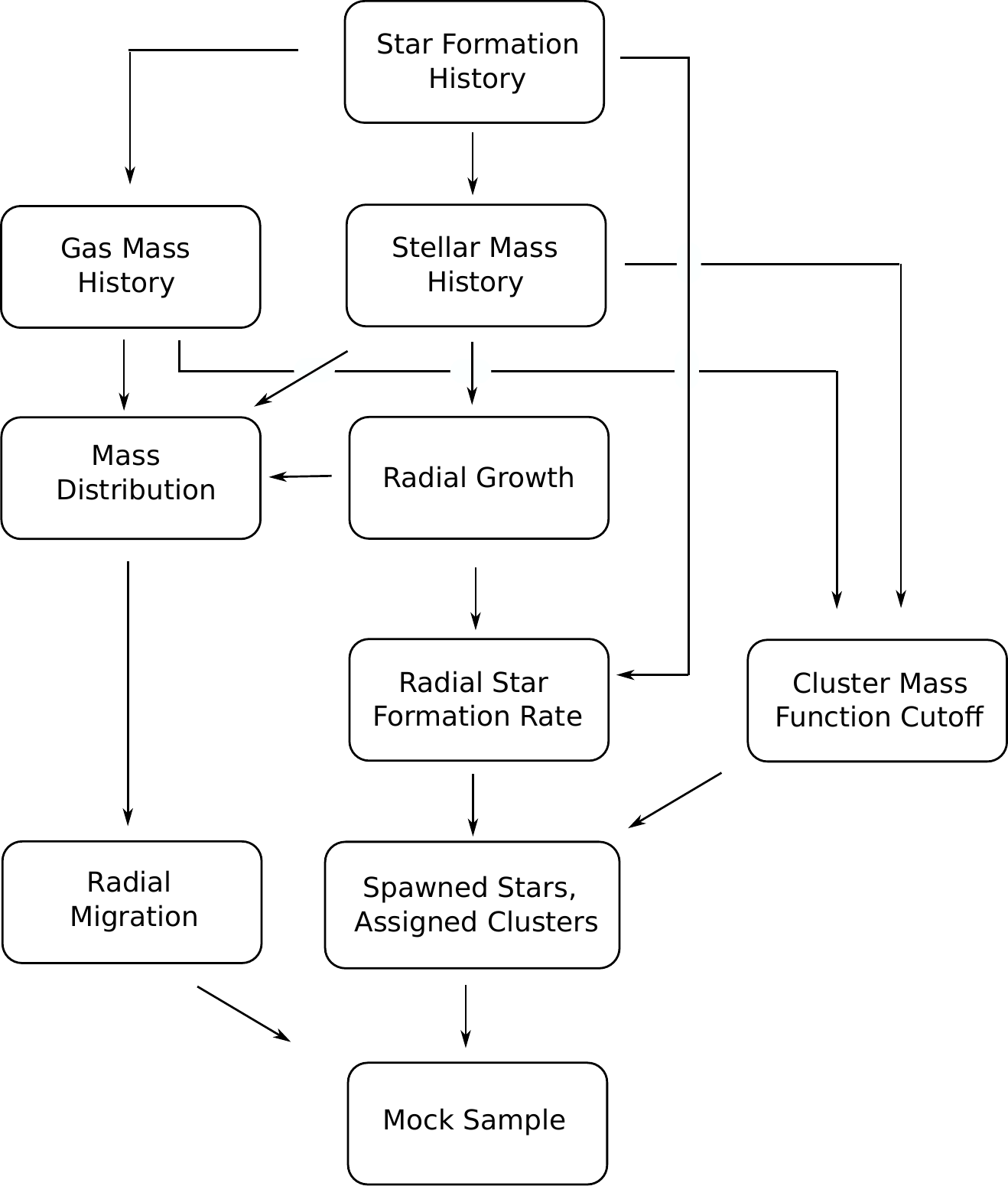}
\caption{Sampling algorithm to create a mock Milky Way data set in this study.}
\label{fig:sampling}
\end{figure}

The sampling algorithm used to create a mock samples is illustrated in Figure \ref{fig:sampling}. To summarize, given a SFH, we obtain the stellar mass evolution through the stellar population synthesis code and the gas mass evolution through the inverted Kennicutt-Schmidt relation. The radial size growth is calculated using an observationally estimated mass-radius relation, which we use to predict the evolution of the SFR scale length. After we obtain the SFR scale length, we calculate the SFR at different radii and different cosmic times from the SFH. We spawn stars through cosmic time according to the radial SFR in discrete time bins of 0.1 Gyr. We only trace stars with $0.5 - 1.5 \, M_\odot$, and we assume a Kroupa IMF.

The gas and stellar masses yield the total mass distribution at different radii and cosmic time. The mass distribution controls the radial migration prescription. The mass distribution is also employed to evaluate the high mass end of the CMF. The CMF is then used to assign a cluster tag to each spawned star, and the radial migration prescription is adopted to mix stars from their birth radii. Note that, we only assign cluster tags after spawning stars in each time bin. We do not generate stars recursively from the CMF although they are both equivalent. In the former case, we avoid a recursive loop in the algorithm and therefore create the mock sample more efficiently. Finally, a mock sample that is within the Solar annulus, given a fixed survey width, is saved for analysis.

Even though the sampling algorithm is straightforward, the effect of radial migration requires us to spawn stars at all radii in the Milky Way disk. In addition, we need to follow each individual star. Therefore, for each set of parameters, we spawn $\sim 10^{11}$ stars, which is computationally expensive even for a semi-analytic model. Each parameter set takes a full CPU day and 50 GB of memory per CPU to evaluate. We evaluate a grid of $\sim 600$ different model parameters. It therefore took $\sim 2$ CPU years to generate the mock samples. After the mock samples were created, we performed Monte Carlo simulations, distributing them into chemical cells. The Monte Carlo simulations required about the same amount of CPU time. Hence, it took $\sim 4$ CPU years in total to generate the results in this study. Including a significant amount of exploratory work, this project consumed $\sim 40$ CPU years of compute time. Obviously, parallelization reduced the total time from $\mathcal{O}$(graduate student lifetime) to $\mathcal{O}$(graduate student year).

\begin{figure}
\center
\includegraphics[width=\textwidth]{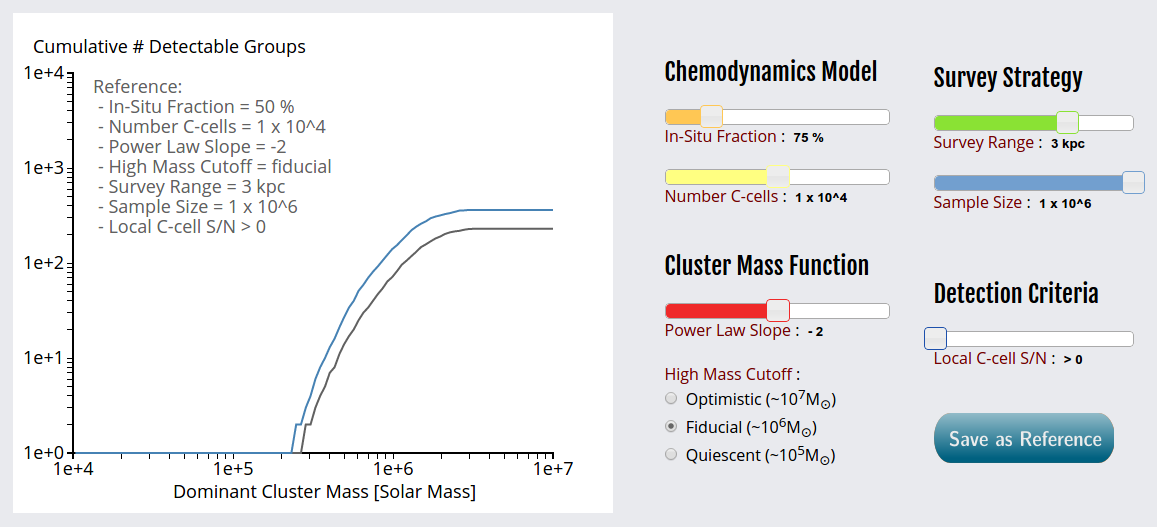}
\caption{A demonstration of the online-applet created in the course of this project.}
\label{fig:online-applet}
\end{figure}

%
%
%
%
%
%

\section{Interactive applet}
\label{sec:interactive}

Since we study a large multidimensional grid of simulations, it is challenging to include all results in this paper. We created an online applet to demonstrate results in the multidimensional grid. In the online applet (\href{www.cfa.harvard.edu/~yuan-sen.ting/chemical_tagging.html}{www.cfa.harvard.edu/$\sim$yuan-sen.ting/chemical$\_$tagging.html}) as shown in Figure~\ref{fig:online-applet}, we plot the cumulative number of detectable groups (exceeding $5\sigma$) as a function of the zero age mass of the dominant cluster. In each detected group, star cluster with the most stars sampled is considered as the dominant cluster.

The applet allows users to change: the in-situ fraction, $f_{\rm in-situ}$, (i.e., the radial migration prescription); the number of chemical cells, $N_{\rm cells}$; the CMF cutoff, $M_{\rm cluster}^{\rm max}$, and slope, $\alpha$; the survey depth, $\Delta R_{\rm survey}$; and the number of stars in the survey, $N_\star$. As demonstrated in Section~\ref{subsec:not-individual}, these detected groups do not necessarily comprise of co-natal stars. The online applet also allows users to impose a local S/N selection criteria as defined in Section~\ref{subsec:not-individual}. For instance, by imposing the criteria local S/N > 1, we select detectable groups that have more stars contributed by the dominant cluster over the combined background from smaller clusters. In the case where no local S/N criteria is imposed, the end point of the cumulative distribution in the applet corresponds to the results in Figure~\ref{fig:local-results-2} and ~\ref{fig:local-results-3}. Finally, there is a ``save as reference'' button in the applet which allows users to save the current cumulative distribution as a reference and compare with the other choices of parameters

\vspace{0.3cm}

\end{CJK*}

\bibliography{biblio}

\end{document}